\documentclass[12pt,hidelinks]{article}
\usepackage[body={17.5cm, 22cm},right=2cm]{geometry}
\usepackage[utf8]{inputenc}
\usepackage{color}
\usepackage{amssymb,graphicx}
\usepackage{amsmath}
\usepackage{epsfig}
\usepackage{lmodern}
\usepackage[normalem]{ulem}
\usepackage{bm}
\usepackage[bookmarks=true,pdfborder={0 0 0}]{hyperref} 
\usepackage[
    natbib=true,
    style=numeric-comp,
    sorting=none
]{biblatex}
\usepackage{booktabs}   
\usepackage{siunitx}    
\usepackage{orcidlink}
\usepackage{graphicx}   
\usepackage{subcaption}
\usepackage{float}
\allowdisplaybreaks

\DeclareUnicodeCharacter{223C}{\ensuremath{\sim}}   
\DeclareUnicodeCharacter{2212}{\ensuremath{-}}      
\DeclareUnicodeCharacter{2243}{\ensuremath{\simeq}} 
\DeclareUnicodeCharacter{2248}{\ensuremath{\approx}}
\DeclareUnicodeCharacter{00B1}{\ensuremath{\pm}}    
\DeclareUnicodeCharacter{2013}{--}                  
\DeclareUnicodeCharacter{2014}{---}    
\DeclareUnicodeCharacter{0308}{\"{}}

\newcommand{\Mpc}{{\, {\rm Mpc}}}

\newcommand{\eV}{{\, {\rm eV}}}
\newcommand{\keV}{{\, {\rm keV}}}

\newcommand{\GeV}{{\, {\rm GeV}}}

\definecolor{myblue}{RGB}{35,20,180}

\begin{document}
\setcounter{page}{0}
\thispagestyle{empty}

\parskip 3pt

\font\mini=cmr10 at 2pt

\begin{titlepage}
\noindent \makebox[11.5cm][l]{\footnotesize \hspace*{-.2cm} }{\footnotesize 
DESY-25-147, CERN-TH-2025-193}  \\  [-1mm]
~\vspace{1cm}
\begin{center}

{\LARGE \bf Large-Scale Structure Probes of}
\\ \vspace{1.6mm}
{\LARGE \bf the Post-Inflationary Axiverse}

\vspace{0.6cm}

		{\large
        Marco~Gorghetto\,\orcidlink{0000-0002-5479-485X},$^a$\,
	Sokratis~Trifinopoulos\,\orcidlink{0000-0002-0492-1144},$^{b,c}$\,
        Georgios~Valogiannis\,\orcidlink{0000-0003-0805-1470}$^{d,e}$}

		\vspace{.6cm}

		{\normalsize { \sl $^{a}$ 
				Deutsches Elektronen-Synchrotron DESY, Notkestr. 85, 22607 Hamburg, Germany}}
            
            \vspace{.2cm}
		{\normalsize { \sl $^{b}$ 
				Theoretical Physics Department, CERN, Geneva, Switzerland}}

            \vspace{.2cm}
		{\normalsize { \sl $^{c}$ 
				Physik-Institut, Universit\"at Z\"urich, 8057 Z\"urich, Switzerland}}
            
            \vspace{.2cm}
		{\normalsize { \sl $^{d}$ 
				Department of Astronomy \& Astrophysics, University of Chicago, Chicago, IL, 60637, USA}}
                
            \vspace{.2cm}
		{\normalsize { \sl $^{e}$ 
				Kavli Institute for Cosmological Physics, Chicago, IL, 60637, USA}}

\vspace{.3cm}

\end{center}

\vspace{1.0cm}
\begin{abstract}

We study the cosmology of axions in the post-inflationary 
scenario, where random initial conditions and the ensuing string–domain-wall network generate an isocurvature power spectrum. Axions radiated from strings behave as warm, wave-like dark matter: when they constitute the full dark matter abundance, free streaming sets the strongest bounds on their mass. For subdominant fractions, despite being warm, they still lead to an overall enhancement of structure growth in the dominant component, seeded by the axion white-noise fluctuations. We search for this effect using the ultraviolet luminosity function (UVLF) of galaxies at $z=4$–$10$, probing $k\simeq0.5$–$10\,\mathrm{Mpc}^{-1}$. Combining the UVLF analysis with Lyman-$\alpha$ and CMB data yields the leading cosmological limits on post-inflationary axion dark matter, sensitive to tiny fractions for $m_a\lesssim10^{-21}\,\mathrm{eV}$. As a byproduct, we obtain new constraints on generic white-noise power spectra from the UVLF. These results apply broadly to scenarios that generate similar isocurvature perturbations, linking early-universe field dynamics to high-redshift structure formation.

\end{abstract}

\end{titlepage}

{\fontsize{11}{10.8}
\tableofcontents
}

\section{Introduction} \label{sec:intro}

Axions are a well-motivated class of Standard Model (SM) extensions. They arise as pseudo-Nambu--Goldstone bosons of spontaneously broken approximate global $U(1)$  symmetries, 
with a shift symmetry protecting their lightness. 
The QCD axion, originally proposed as a dynamical solution to the strong CP problem~\cite{Peccei:1977hh,Peccei:1977ur,Weinberg:1977ma,Wilczek:1977pj}, stands out as a robust dark matter  candidate~\cite{Abbott:1982af,Dine:1982ah,Preskill:1982cy,Marsh:2015xka,Adams:2022pbo}. 
More general axions are produced automatically as dark matter relics and appear  in ultraviolet  completions such as string theory~\cite{Svrcek:2006yi,Arvanitaki:2009fg,Petrossian-Byrne:2025mto}, 
where they often arise as the pseudoscalar partners of moduli fields. String models realize the QCD axion while also producing multiple ultralight axions — a string `axiverse'. For reviews on axions, see e.g.~\cite{Marsh:2015xka,Graham:2015ouw,Hook:2018dlk,Irastorza:2018dyq,Agrawal:2021dbo}. 

Whether the $U(1)$ symmetry is broken before or after inflation has important cosmological consequences. 
If it is broken before inflation, the axion field takes a homogeneous value across our observable Universe, and its relic abundance is set by the misalignment mechanism~\cite{Preskill:1982cy,Abbott:1982af,Dine:1982ah}. 
If it is instead  broken after inflation, the axion acquires uncorrelated initial conditions in different regions of the Universe, leading to the formation of topological defects such as cosmic strings and domain walls~\cite{Sikivie:1982qv,Vilenkin:1982ks,Vilenkin:1984ib,
Davis:1986xc}. 
The decay of these nonlinear objects contributes to, and likely dominates, the axion dark matter relic abundance. It also imprints large spatial inhomogeneities in the dark matter density field at comoving scales $k_\star$ set by the Hubble parameter at the time of their decay, which is of order the axion mass. For the QCD axion (whose mass is relatively large, $m_a \gtrsim 0.03-0.5\,\text{meV}$,  to avoid dark matter overproduction~\cite{Gorghetto:2020qws,Saikawa:2024bta,Buschmann:2021sdq,Benabou:2024msj,Kim:2024wku}), this corresponds to $k_\star^{-1} \ll 1\,\text{Mpc}$, i.e. well below galactic scales. However, for sufficiently light axions -- focus of this work  -- the corresponding spatial scales are substantially larger, e.g. of ${\cal O}(\text{Mpc})$ for $m_a\lesssim 10^{-20}$\,eV.

The axion dark matter  inhomogeneities are isocurvature fluctuations and affect structure formation in multiple ways. 
First, they enhance the matter power spectrum via a tail of  white-noise perturbations at scales $k\lesssim k_\star$~\cite{Kobayashi:2017jcf,Irsic:2019iff,Feix:2019lpo,Feix:2020txt,Gorghetto:2022ikz,Hutsi:2022fzw,Bird:2023pkr}. These evolve as cold dark matter (CDM) on large spatial scales ($k \ll k_\star$), increasing the number of dark matter halos and observable galaxies. Such structure formation is instead inhibited on smaller scales. Indeed, the fluctuations correspond to the axions being produced as \emph{warm} wave-like dark matter, as they are semi-relativistic at production but light enough 
to have large occupation number when they subsequently turn nonrelativistic. Consequently, the axion fluctuations \emph{do not} grow at spatial scales below the Jeans scales associated  with either the warm dark matter velocity dispersion or its quantum pressure, with both Jeans scales lying close to $k_\star$.  

In addition, the warm axions free-stream over cosmological distances, erasing existing correlations between the adiabatic modes and effectively suppressing the adiabatic power spectrum below the so-called free-streaming length~\cite{Amin:2022nlh,Ling:2024qfv,Liu:2024pjg,Amin:2025sla,Amin:2025dtd}, which can lie at observable scales. This effect provides the strongest constraint on light post-inflationary axions, restricting their mass to $m_a\gtrsim10^{-18}$~eV if they constitute  a sizable fraction of the dark matter~\cite{Amin:2022nlh}.  
In this work we thus focus on axion masses $m_a$ and decay constants $f_a$ such that the axion constitutes a subcomponent of dark matter (approximately $m_a \lesssim 10^{-20}\,\text{eV}$ and $f_a \lesssim 10^{15}\,\text{GeV}$), and assume that the dominant component behaves as CDM, with only adiabatic perturbations initially. We demonstrate that in this regime the free-streaming suppression is reduced, and the corresponding bounds on $m_a$ are relaxed. Crucially, even though the axion constitutes only a dark matter subcomponent, the isocurvature tail can enhance structure formation, including on scales where the axion dark matter growth is suppressed by its velocity dispersion or quantum pressure. This is because, although the axion dark matter has a warm wave-like behavior at momenta of order (and slightly below) $k_\star$, its isocurvature fluctuations rapidly induce similar isocurvature fluctuations in the dominant CDM component, which can increase and eventually collapse. 
 
This enhanced dark matter structure lies squarely within the reach of large-scale structure (LSS) observations, which trace the growth of 
primordial fluctuations into the web of galaxies and matter observed today~\cite{Frenk:2012ph}. LSS observations such as galaxy clustering~\cite{BOSS:2016wmc,2010MNRAS.404...60R,DESI:2025zgx}, weak gravitational lensing~\cite{DES:2021wwk}, stellar streams~\cite{Banik:2019smi} and cosmic microwave background (CMB) measurements~\cite{Planck:2018nkj,Planck:2018vyg,CMB-S4:2016ple} constrain the matter power spectrum 
as predicted by $\Lambda$CDM at the percent level on cosmological scales ($k\lesssim\,\text{Mpc}^{-1}$), but lose sensitivity at wavenumbers above $0.5\,\text{Mpc}^{-1}$. In contrast, Lyman-$\alpha$ observations~\cite{eBOSS:2018qyj,Chabanier:2019eai,Irsic:2017yje} as well as current and future data from ultraviolet luminosity functions (UVLFs) of galaxies can extend the cosmological constraints to higher redshifts and much smaller scales up to $\sim 10\,\text{Mpc}^{-1}$. In particular, UVLFs probe the comoving number density of galaxies as a function of their ultraviolet brightness redshift, and at present they offer access to the highest redshifts among other LSS observations, with Hubble Space Telescope (HST) data reaching $z=4\text{--}10$~\cite{Bouwens:2014fua,2015ApJ...810...71F,Atek:2015axa,Livermore:2016mbs,2017ApJ...843..129B,2017ApJ...838...29M,2018ApJ...854...73I,2018ApJ...855..105O,Atek:2018nsc,2020ApJ...891..146R,2021AJ....162...47B} and recent James Webb  Space Telescope (JWST) observations pushing this frontier to $z\sim25$~\cite{2022ApJ...940L..14N,Castellano:2022ikm,2023MNRAS.518.6011D,2023ApJ...946L..35M,2023MNRAS.523.1036B,2023ApJ...951L...1P,2023ApJS..265....5H,2024ApJ...960...56H,Perez-Gonzalez:2025bqr,Castellano:2025vkt}. LSS observations have proven to be a powerful tool in constraining scenarios of new physics that induce a modification of the matter power spectrum~\cite{Kobayashi:2017jcf,Irsic:2019iff,Feix:2019lpo,Feix:2020txt,Co:2021lkc,Hutsi:2022fzw,Gorghetto:2022ikz,Bird:2023pkr,Winch:2024mrt,Ellis:2025xju,Afshordi:2003zb,Carr:2018rid,Inman:2019wvr,Murgia:2019duy,Liu:2022bvr,Gouttenoire:2023nzr,Buckley:2025zgh,Khan:2025kag,Gerlach:2025vco,Co:2025lrd,Chatrchyan:2023cmz,Eroncel:2022efc}.

In this work, we present a comprehensive LSS analysis of post-inflationary axions, introducing two key advances. First, we model  the axion-induced matter power spectrum using results from recent field-theory simulations of axion string networks~\cite{Gorghetto:2021fsn,Gorghetto:2022ikz}. These indicate that axion inhomogeneities peak at comoving scales of $\mathcal{O}(10)\,k_\star$ rather than $k_\star$, reducing the predicted isocurvature amplitude for a fixed $m_a$ and enhancing the velocity-dispersion effects, as already noted in Ref.~\cite{Petrossian-Byrne:2025mto}. This suppresses axion dark matter perturbation growth below the corresponding Jeans scale, but simultaneously triggers the growth of structures at momenta $\mathcal{O}(10)k_\star$ in the dominant cold component. 
Second, we derive new observational constraints on white-noise–type power spectra using the ultraviolet luminosity function (UVLF) of galaxies observed by the Hubble Space Telescope (HST), and apply these bounds self-consistently to the axion+CDM system. Our UVLF analysis employs the \texttt{GALLUMI} likelihood framework~\cite{Sabti:2021unj,Sabti:2021xvh}, which marginalizes over astrophysical uncertainties—such as star-formation efficiency and the halo–galaxy connection—and is based on physically motivated priors calibrated to $\Lambda$CDM fits.\footnote{Refs.~\cite{Winch:2024mrt,Ellis:2025xju} have employed similar methodology to study warm and fuzzy dark matter, which exhibit suppression of the matter power spectrum instead of enhancement as in our case.}  

Finally, we adapt to the post-inflationary axion scenario the bounds derived in Refs.~\cite{Feix:2020txt,Ivanov:2025pbu,Buckley:2025zgh} on  generic white-noise spectra  and primordial black holes from Lyman-$\alpha$, CMB, and BAO observations. Taken together, the bounds in our work provide the strongest constraints when the axion constitutes a dark matter subdominant component.

The article is structured as follows. In Sec.~\ref{sec:theory} we review the post-inflationary scenario, focusing on the generation of inhomogeneities from string and domain wall decay and their imprint on the matter power spectrum. We also analyze the axion warm-dark-matter effects -- namely, free streaming and Jeans suppression driven by velocity dispersion -- on the CDM component, and how its fluctuations grow. Sec.~\ref{sec:LSS} explores the response of large-scale structure observables to this component, with emphasis on the novel UVLF analysis. The resulting constraints are collected in Fig.~\ref{fig:Piso}, and their impact on the total matter power spectrum is shown in Fig.~\ref{fig:Ptot}. In Sec. \ref{sec:results}, we present the resulting bounds on the axion mass and decay constant, providing a summary of the current state-of-the-art exclusion limits in Fig.~\ref{fig:money_plot}. Finally, Sec.~\ref{sec:conclusions} provides a summary of our findings and discusses future directions.

\section{Axion imprints on the power spectrum}
\label{sec:theory}

\subsection{Post-inflationary axion cosmology
}
 
\begin{figure}[t]
    \centering
        \includegraphics[width=0.482\textwidth]{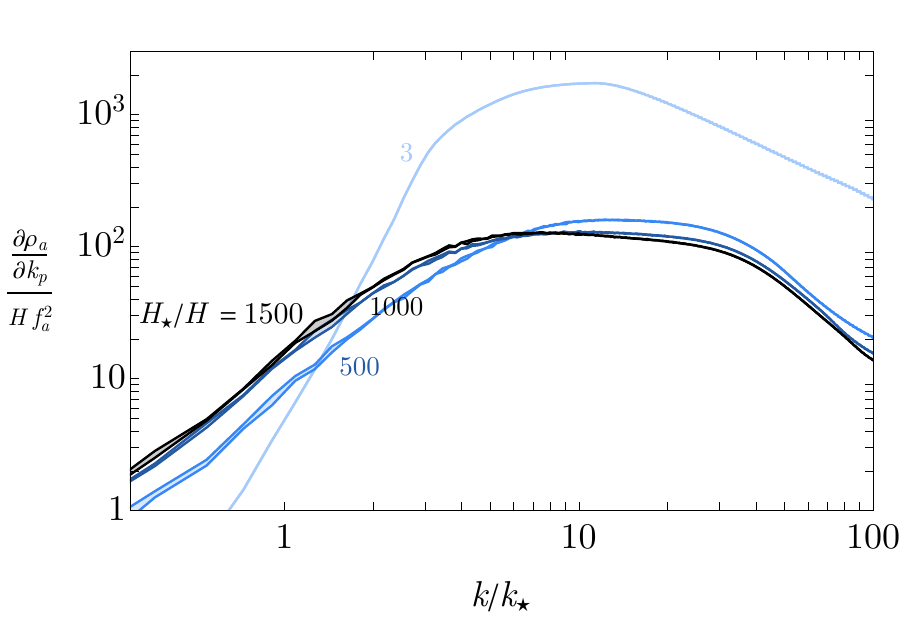} \ \ \ 
        \includegraphics[width=0.47\textwidth]{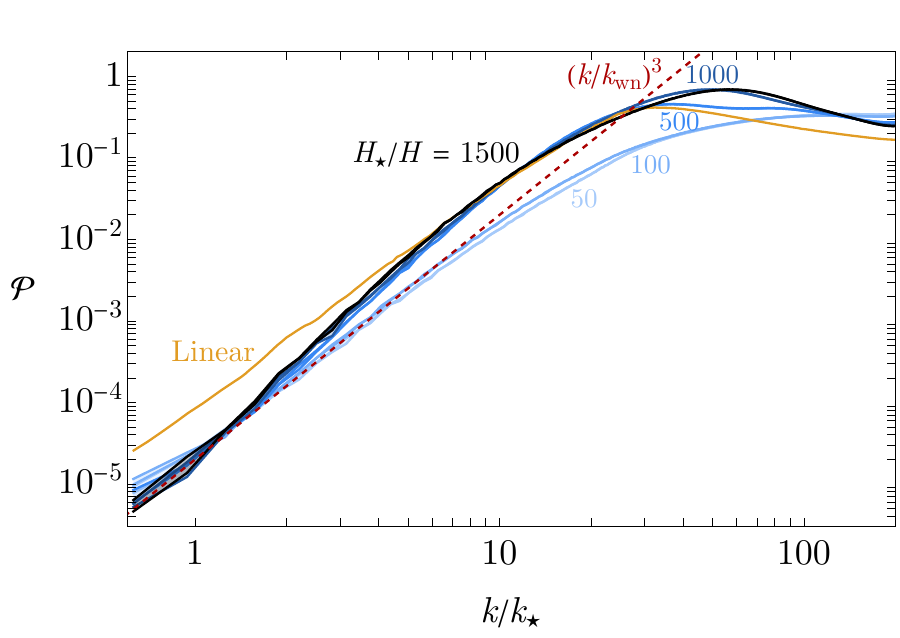} \vspace{-1mm}
     \caption{{\small The energy density spectrum (left) and the overdensity power spectrum (right) of the axion dark matter as it is expected to be produced by axion strings at $\log(m_r/H)\gg1$, as a function of the comoving momentum $k$ normalized to $k_\star = m_a a_\star$; $k_p=k/a$ is the physical momentum. The blue lines show the nontrivial  evolution of the spectrum around $H = H_\ell \simeq H_\star/200$, when the axions emitted from strings experience a nonlinear transient as the axion potential becomes relevant. This increases mildly the typical momentum scale where $\mathcal{P}$ peaks, with respect to a purely linear evolution (orange line). At the end of the nonlinear transient (black line), $\mathcal{P}$ is of order one at $k_{\rm wn}=Ck_\star$ with $C=\mathcal{O}(10)$ or slightly larger. At $k\ll k_{\rm wn}$, it acquires the white-noise form $(k/k_{\rm wn})^3$.
     }}
     \label{fig:Piso-sim}
\end{figure}

We study the cosmology of a theory with a global $U(1)$ Peccei--Quinn (PQ) 
symmetry, that is spontaneously broken by the vacuum expectation value $v$ of a complex scalar field $\varphi$, with $v$ lower than the inflationary Hubble scale or the reheating temperature. Expanding $\phi$ around the vacuum, 
\begin{equation}
    \varphi(x) = \frac{1}{\sqrt{2}}\,(r(x)+v)\,e^{i\phi(x)/v}\,,
\end{equation}
the field $\varphi$ decomposes into a heavy radial excitation $r$ and an angular degree of freedom $\phi$, which is identified as the axion. 
 The dynamics of $\varphi$ are governed by the potential
\begin{equation}\label{eq:potential}
    V_\varphi(\varphi) = \frac{m_r^2}{2v^2}\left(|\varphi|^2 - \tfrac{v^2}{2}\right)^2\, ,
\end{equation}
where $m_r \simeq v$ is the mass of the radial mode. We assume the axion has a temperature-independent potential $V(\phi)$ (to be added to Eq.\,\eqref{eq:potential}) periodic of $2\pi f_a$, where $f_a$ is the axion decay constant, and leading to the temperature-independent mass $m_a$.\footnote{The simulation in Fig.~\ref{fig:Piso-sim} uses $V(\phi) = -m_a^2 f_a^2 \cos(\phi / f_a)$, but its precise form is not relevant for this discussion.} The axion constitutes a fraction $f_{\rm DM}$ of the dark matter, such that its relic density is $\Omega_a = f_{\rm DM}\,\Omega_{\rm DM}$.\footnote{It is customary in the particle physics literature to reserve the term \emph{axion} for the QCD case with a temperature-dependent mass, while temperature-independent scenarios are often called \emph{axion-like particles}. In this work, we focus on the cosmology of such particles and, for brevity, will refer to the particle under study simply as an axion, unless explicitly stated otherwise for the QCD axion.} Interactions with the thermal plasma are negligible at temperatures $T\ll v$, and the dynamics are described entirely by Eq.\,\eqref{eq:potential}. After the PQ phase transition, the axion field $\phi$ acquires random values in its fundamental domain, $\phi/v \in (-\pi,\pi)$, on patches smaller than the horizon. This induces the formation of a network of axion cosmic strings through the Kibble mechanism~\cite{Kibble:1976sj,Vilenkin:1981kz,Vilenkin:1982ks,Sikivie:1982qv,Sikivie:2006ni}, around which $\phi/v$ winds by $2\pi$. 

As the Universe expands, the string network evolves toward a scaling regime in which the number of long strings per Hubble volume, denoted by $\xi$, remains approximately constant. In this regime, the string energy density is $\rho^{\rm st} \simeq 4\xi \mu H^2$, where $H=\dot{a}/a$ is the Hubble parameter and $\mu \simeq \pi v^2 \log(m_r/H)$ is the string tension. Numerical simulations show that $\xi(t) \simeq 0.24 \log(m_r/H)$ increases only logarithmically with time~\cite{Gorghetto:2018myk,Gorghetto:2021fsn,Gorghetto:2022ikz}. To sustain the scaling solution, the network continuously radiates its energy into (initially relativistic) axions at a rate $\Gamma^{\rm st} \simeq 2 \rho^{\rm st} H$, which subsequently redshift and contribute to the present-day dark matter abundance and to dark radiation.

The scaling regime persists until the axion potential $V(\phi)$ becomes cosmologically relevant, at time $t_\star$ defined as $H(t_\star)\equiv H_\star = m_a$ (in what follows, quantities with index $\star$ are evaluated at $t=t_\star$). The axion potential 
breaks the PQ symmetry down to a $\mathbb{Z}_N$ subgroup, where $N \equiv v / f_a \in \mathbb{Z}$. At $t_\star$, $N$ domain walls appear attached to each string, interpolating between two of the $N$ distinct minima of $V(\phi)$ wound around the string~\cite{Sikivie:1982qv,Zeldovich:1974uw,Gelmini:1988sf,Larsson:1996sp,Gelmini:2021yzu}.  The evolution at $H<m_a$ depends on $N$. For $N = 1$, the case primarily considered in this work, each string is attached to a single domain wall that terminates on another string with opposite axion winding. The wall tension then pulls strings and anti-strings together, causing the string-wall system to collapse in one Hubble time. 
The energy stored in the string-wall network is released as additional axions, contributing to the relic abundance alongside those produced during the earlier scaling regime. In the following discussion, and throughout the rest of this paper, we will assume $N=1$, while the more involved case $N>1$ will be addressed in a dedicated paragraph at the end of this subsection.

\vspace{2mm}
\noindent
{\bf Dark matter abundance.}  

\noindent 
Once the string network has formed, the subsequent cosmological evolution of the axion is governed primarily by two parameters: $m_a$ and $f_a$. 
The relic abundance is fixed as a function of these, and, as mentioned, receives contributions from the string scaling regime and from the eventual string–wall decay.\footnote{For the QCD axion, the relic density is also affected by the temperature-dependence of the axion mass induced by nonperturbative QCD dynamics.}
Sizable theoretical uncertainties remain in quantifying both contributions due to the highly nonlinear string and domain-wall dynamics. 

Focusing first on the scaling regime, two key ingredients control the relic abundance: the string density $\xi(t)$ and the spectrum of axions $\partial\Gamma^{\rm st}/\partial k$ radiated by the network. 
Both must be extracted from numerical simulations, which are restricted to unphysical values $\log(m_r/H)\lesssim 8-10$ ~\cite{Gorghetto:2018myk,Gorghetto:2020qws,Saikawa:2024bta,Kim:2024wku,Buschmann:2021sdq,Benabou:2024msj}, and therefore require extrapolation to the cosmological regime where $\log(m_r/H)\simeq 100$. 
If the axion emission is dominated by infrared (IR) momenta -- i.e. if $\partial\Gamma^{\rm st}/\partial k$ peaks at momenta of order Hubble, $k/a=x_0H$ with $x_0=\mathcal{O}(10)$, as suggested by simulations~\cite{Gorghetto:2020qws} -- then the axion number density is controlled by the latest stages of the scaling regime. 
At $t_\star$ this yields the energy density (at IR momenta) $\rho_\star^{\rm st}\simeq\Gamma_\star^{\rm st}/H_\star\simeq 8\pi f_a^2 \xi_\star \log_\star H_\star^2$ and the corresponding number density
\begin{equation}
n_{\star}^{\rm st} \simeq \frac{\rho_\star^{\rm st}}{x_0 H_\star}
   \simeq \frac{8\pi f_a^2 \xi_\star \log_\star}{x_0} H_\star \,,
\end{equation}
where $\log_\star=\log(m_r/H_\star)\simeq \log(f_a/m_a)$. 
At $t>t_\star$ the axions become non-relativistic and their comoving number density is subsequently conserved, leading to a present-day relic density
\begin{equation}
    \label{eq:relic}
\Omega_a^{\rm st} \equiv \frac{\rho_0^{\rm st}}{\rho_{\rm crit}}  \simeq 0.081 \left(\frac{\xi_\star\log_\star}{2.4\cdot 10^3} \right) \left(\frac{f_a}{10^{14}\,{\rm GeV}}\right)^{2}\left( \frac{m_a}{10^{-18}\,\eV} \right)^{1/2}  \left(\frac{10}{x_0}\right) \left(\frac{3.5}{g_\star (T_\star)} \right)^{1/4}\,,
\end{equation}
where $g_\star$ the number of relativistic species in equilibrium and we used $\rho_0^{\rm st}=m_an_0^{\rm st}=m_a(a_{\star}/a_0)^3n_{\star}^{\rm st}$. 
For $m_a\in [10^{-28},10^{-18}]\,\eV$ and $f_a\in[10^{14},10^{15}]\,\GeV$ where LSS probes are relevant, we have $\log_\star\simeq 100$ and $\xi_\star\log_\star\simeq 2.4 \cdot 10^3$, assuming the logarithmic growth of $\xi$ persists at $\log(m_r/H)\gg 1$.

Eq.\,\eqref{eq:relic} holds at best up to an $\mathcal{O}(1)$ factor, which captures the effect of the string-wall system on the axions emitted during the scaling regime. 
Additionally, for $\xi_\star\log_\star\gtrsim 10^3$, because the axion kinetic energy ($\sim\rho_\star^{\rm st}$) is much larger than the potential energy $\sim m_a^2f_a^2$ at $t_\star$, the axions keep redshifting relativistically after $t_\star$ (with energy $\sim\rho_\star^{\rm st}(a_\star/a)^4$), until two match, at $H_\ell\simeq m_a/(4\pi \xi_\star\log_\star)^{1/2}$. This leads to a suppression of the abundance $\Omega_a^{\rm st}\propto (\xi_\star\log_\star)^{3/4}$, as well as to a nonlinear transient that boosts the typical axion comoving momentum to $k\simeq (4\pi \xi_\star\log_\star)^{1/4} m_aa_\star$~\cite{Gorghetto:2020qws}. While this effect is crucial for the QCD axion due to its temperature-dependent mass, it amounts to about 20\% for an axion with a temperature-independent mass -- comparable to other uncertainties in Eq.\,\eqref{eq:relic} -- and is therefore neglected in Eq.\,\eqref{eq:relic}~\cite{Gorghetto:2021fsn,Gorghetto:2022ikz}. Finally, the axion emission during the scaling regime might occur at momenta parametrically higher than $H$, e.g. with a scale-invariant spectrum, as suggested in~\cite{Buschmann:2021sdq,Benabou:2024msj,Hagmann:1990mj}, in which case $\Omega_a^{\rm st}$ is suppressed by a factor of $\log_\star$ or more.

The contribution from the string-wall decay at $t=t_\star$ is even more uncertain. Let $\mathcal A$ be the number of domain walls per Hubble patch, and $\sigma=\beta m_a f_a^2$ the wall surface tension,  with $\beta\simeq 8$ for a cosine-like axion potential.
The walls emit axions with energy density $\rho^{\rm w} = \mathcal{A} \sigma H$ and the rate $\Gamma^{\rm w} \simeq  2 \rho^{\rm w}  H$. If they collapse shortly after $t_\star$, their energy is primarily transferred to axions with momentum $\sim H_\star$, yielding an energy density $\rho_{\rm w}\sim \Gamma^{\rm w}_\star/H_\star$ and a number density 
\begin{equation}
    n_{\star}^{\rm w} \simeq \frac{\rho_{\rm w}}{H_\star} \simeq 2\beta \xi_\star m_a f_a^2\,,
\end{equation}
where we used $\mathcal{A} \simeq 0.35\log(m_r/H) \simeq \xi$ during scaling (see~\cite{Gorghetto:2022ikz}).  
This suggests that $n_{\star}^{\rm w}$ may be smaller than the string contribution $n^{\rm st}_\star$ in Eq.\,\eqref{eq:relic} by an $\mathcal{O}(10)$ factor, as $n^{\rm w}_\star$ is not enhanced by $\log_\star$. However, this may underestimate the axion abundance from domain walls if their decay occurs parametrically later than $t_\star$.  
Given these uncertainties, throughout this work we adopt Eq.~\eqref{eq:relic} as the estimate of the total axion abundance 
\begin{equation}\label{eq:Omegast}
    \Omega_a = \Omega_a^{\rm st}\,,
\end{equation}
with the understanding that this represents a conservative lower bound.

\vspace{3mm}
\noindent
{\bf Dark matter power spectrum and isocurvature component.}  

\noindent
The axion dark matter field resulting from string and domain-wall decay develops $\mathcal{O}(1)$ inhomogeneities on length scales roughly set by the mean string separation, parametrically of order the Hubble parameter $H_\star=m_a$. These correspond to isocurvature fluctuations, since they reside entirely in the axion. We write the total dark matter overdensity field as  
\begin{equation}\label{eq:delta}
    \delta = f_{\rm DM}\,\delta_a + (1-f_{\rm DM})\,\delta_c \,, \quad \delta_a=\frac{\rho_a-\bar{\rho}_a}{\bar{\rho}_a} \,, \quad \delta_c=\frac{\rho_c-\bar{\rho}_c}{\bar{\rho}_c} \,,
\end{equation}
where ${\rho}_a$ and ${\rho}_c$ are the axion and additional dark matter component energy densities, respectively, the bar denotes their average, and $f_{\rm DM}=\bar{\rho}_a/(\bar{\rho}_a+\bar{\rho}_c)$. The fluctuations of the overdensity field $\delta$ are described by the dimensionless power spectrum $\mathcal{P}(k)$ defined by
\begin{equation}\label{eq:powerspec}
    \langle \delta^*(k)\,\delta(k')\rangle 
      = \frac{2\pi^2}{k^3}\,(2\pi)^3\delta^3(k-k')\,\mathcal{P}(k)\,, 
\end{equation}
where $\delta(k)=\int d^3x\,e^{-ikx}\,\delta(x)$ is the Fourier transform  of $\delta(x)$.

The $\mathcal{O}(1)$ axion overdensity fluctuations are seeded at comoving momenta $C k_\star$, near the peak $k_a$ of the axion energy density spectrum $\partial\rho_a/\partial k$,\footnote{This is defined by $\rho_a=\int dk \partial\rho_a/\partial k$. In the non-relativistic limit, when $H\ll H_\ell$, $\rho_a\simeq \frac12 (\dot{a}^2+ \nabla^2 a+m_a^2a^2)$.} with
\begin{equation}\label{eq:kstar}
    k_\star \equiv m_a a_\star = H_\star a_\star 
      \simeq 54 \left(\frac{m_a}{10^{-20}\,\eV}\right)^\frac12\Mpc^{-1} \, .
\end{equation}
Simulations  at $\log(m_r/H)\lesssim8$  suggest that $C$ is $\mathcal{O}(10)$~\cite{Gorghetto:2021fsn,Gorghetto:2022ikz}.\footnote{This $\mathcal{O}(10)$ factor likely reflects that the emission occurs at \emph{wavelengths} of order Hubble, $2\pi / (k/a) \simeq H^{-1}$, rather than $k/a \simeq H$. For the calculation of $a_{\star}$ in Eq.~\eqref{eq:kstar} we used $g_{\star,\epsilon}(T_\star)=3.363$ and $g_{\star,s}(T_\star)=3.909$.} 
In addition, after the small nonlinear evolution of the axions emitted by strings occurring at $H=H_\ell$,  the typical  comoving axion momentum is driven to 
\begin{equation}
k \simeq (4\pi \xi_\star \log_\star)^{1/4} m_a a_\star \,=\, \mathcal{O}(10)\,k_\star \, .
\end{equation}
 These results are summarized in Fig.\,\ref{fig:Piso-sim}, adapted from Ref.~\cite{Gorghetto:2021fsn}. The left panel shows the axion energy density spectrum $\partial\rho_a/\partial k$  from a simulation of the axions from the string scaling regime at different times through the nonlinear transient around $H\simeq H_\ell\simeq H_\star/200$, with the spectrum at the end of it (black line) peaked at $k_a\simeq O(10) k_\star$. At late times, the axions redshifts as free nonrelativistic matter, i.e. $\rho_a\propto a^{-3}$, hence the normalization to $H\propto a^{-2}$ in Fig.~\ref{fig:Piso-sim}. Meanwhile, the right panel displays the corresponding overdensity spectrum $\mathcal{P}$, whose peak occurs at $k = C k_\star$, with $C \simeq 10-40$, comparable to that of the energy spectrum.

  If, however, domain-wall decay turns out to provide the dominant contribution to the dark matter abundance, the characteristic momentum peak may be at smaller values, potentially reducing $C$ to $\mathcal{O}(1)$~\cite{Gorghetto:2021fsn}. Note that for $m_a \lesssim 10^{-20}\,\text{eV}$, the characteristic momentum scale $k_\star$ in Eq.~\eqref{eq:kstar} is sufficiently small to be accessible to LSS observations.

At spatial scales larger than the scale
\begin{equation}\label{eq:kwn}
k_{\rm wn}\equiv Ck_\star ,
\end{equation}
the dark matter isocurvature fluctuations are uncorrelated by causality. This implies that this isocurvature component $\mathcal{P}_{\rm iso}$ to the matter power spectrum  assumes a white-noise form, as evident in Fig.~\ref{fig:Piso-sim}. On much larger  scales, consistent with $\Lambda$CDM, the total dark matter field follows the standard spectrum of adiabatic perturbations $\mathcal{P}_{\rm ad}(k)=A_s(k/k_p)^{n_s-1}$ with $A_s\simeq 2.2\cdot 10^{-9}$, $n_s\simeq 0.96$, and $k_p=0.05\,{\rm Mpc}^{-1}$, which dominates over the isocurvature part. As a result, given the large uncertainties on $\mathcal{P}_{\rm iso}$, we conservatively approximate the total matter power spectrum by
\begin{align} \label{eq:Piso}
 \mathcal{P}(k) &= \mathcal{P}_{\rm ad}(k)+\mathcal{P}_{\rm iso}(k)\,, \notag \\ 
    \mathcal{P}_{\rm iso}(k) &=  \begin{cases}
      f_{\rm DM}^2 (k/k_{\rm wn})^3
      ~,& \text{if } k\lesssim  k_{\rm wn} \ \ \\
    0~,              & \text{otherwise}
\end{cases}~,
\end{align}
where $k_{\rm wn}$ is defined in Eq.\,\eqref{eq:kwn} and sets the momentum where the spectrum is of order one for $f_{\rm DM}=1$.  $\mathcal{P}(k)$ in Eq.\,\eqref{eq:Piso} is defined just after the axion dark matter is produced, at $a_\star\ll a_{\rm eq}$, where $a_{\rm eq}$ is the scale factor at matter-radiation equality. Note that $\mathcal{P}_{\rm ad}$ arises from the adiabatic perturbations of both the axion and the additional CDM ($\delta_a$ and $\delta_c$), while $\mathcal{P}_{\rm iso}$ only from the axion (at least at $a\ll a_{\rm eq}$, before the onset of linear growth). 

Given  uncertainties on the axion spectrum, and to streamline the discussion, in the rest of the paper we will not track explicitly the difference between the peak $k_a$ of the axion spectrum and $k_{\rm wn}$ (this is conventional in the literature; see, e.g., Ref.~\cite{Amin:2022nlh}). When presenting bounds, we instead adopt the benchmark $C=10$ and the more conservative choice $C=20$. The difference between these provides an estimate of the uncertainty in $k_{\rm wn}$ and the potential mismatch between $k_a$ and $k_{\rm wn}$.

\vspace{2.5mm}
\noindent
{\bf Domain wall number \boldsymbol{$N > 1$}.}\nopagebreak  

\noindent
For $N>1$, the string–wall network that forms at $t_\star$ is cosmologically stable, since each string is pulled in opposite directions with the same force by multiple domain walls~\cite{Sikivie:1982qv,Zeldovich:1974uw,Gelmini:1988sf,Larsson:1996sp,Gelmini:2021yzu}. 
Similarly to the string network, such a long-lived system is expected to enter a scaling regime, in which the domain wall area per Hubble patch $\mathcal{A}$ remains approximately constant through the continuous emission of axions. In the limit of an exact $\mathbb{Z}_N$ symmetry, this scaling regime persists indefinitely and the wall energy density eventually dominates the Universe, thereby ruling out the theory.

Consequently, small $\mathbb{Z}_N$-breaking potential $\delta V$ is necessary to lift the vacuum degeneracy and select a unique minimum of $V$.\footnote{Notice that the QCD axion requires the PQ-breaking potential to be small enough to satisfy bounds on the induced effective strong CP-violating $\theta_{\rm QCD}$ parameter, whereas for generic axions there is no analogous constraint.} This becomes cosmologically relevant when  $\rho_w \simeq \delta V$, triggering the network's decay, analogous to the $N=1$ case but at a lower Hubble scale $H_d<H_\star$, set by $\delta V$. The domain wall energy density, $\rho_d^{\rm w} = \mathcal{A}_d\sigma H_d$, released in this process is also expected to be converted into axions (quantities with index $d$ are evaluated at $t=t_d$ with $H(t_d)\equiv H_d$). 

Although simulations have not yet confirmed this, it is plausible that the domain-wall decay enhances the relic abundance relative to the string contribution, which for $N>1$ is simply given by Eq.\,\eqref{eq:relic} with the substitution $f_a\to N f_a$. Specifically, if  all the wall energy released at $H_d$ --  comparable to the energy emitted during the wall scaling regime evaluated at $H_d$ -- is converted into axions with momenta of order $m_a$ or less, the resulting relic density is~\cite{Gorghetto:2022ikz,Petrossian-Byrne:2025mto}  

\begin{equation}\label{eq:Omegaw}
    \Omega_{a,N>1}^{\rm w}\simeq 0.026 \mathcal{A}_d\left(\frac{m_a}{H_d}\right)^\frac12\left(\frac{f_a}{10^{12}\GeV}\right)^2\left(\frac{m_a}{10^{-6}\eV}\right)^\frac12 \,.
\end{equation}
In Eq.\,\eqref{eq:Omegaw}, the dependence on $N$ enters through $\mathcal{A}_d$. For $H_d\ll m_a$, i.e. $\delta V$ sufficiently small, the axions with $N>1$ can thus account for the observed dark matter for smaller $f_a$ than the $N=1$ case (partly because the production is delayed until $H = H_d$, rather than occurring at $H \simeq m_a$, resulting in a smaller redshift factor). We take Eq.\,\eqref{eq:Omegaw} as an estimate of the dark matter abundance for $N > 1$.
 
In this scenario it is plausible that the $\mathcal{O}(1)$ fluctuations arise at the comoving momentum scale $k_{\rm dw}$ set by the inverse comoving domain wall size $k_{\rm dw}= C_d a_d H_d $ when they decay, where $C_d$ is unknown but, based on simulations with $N=1$~\cite{Gorghetto:2021fsn}, is possibly smaller than the coefficient discussed earlier in the case of strings. The corresponding comoving momentum scales are thus parametrically smaller than the characteristic scale, $Ck_\star$, expected for $N = 1$.  As suggested by simulations~\cite{Gorghetto:2022ikz}, the axions are likely to be produced semi-relativistic with momentum $k/a_d \sim m_a$, which is much larger than $H_d$, and they rapidly become nonrelativistic as the Universe expands. As we will discuss in Sec.~\ref{sec:results}, structure formation imposes a lower bound on $H_d$, constraining the values of $m_a$ and $f_a$ for which the axion can comprise the dark matter, $\Omega_{a,N>1}^{\rm w}=\Omega_{\rm dm}$.  

\subsection{Free streaming and classical/quantum Jeans scales}
\label{sec:scales}

Shortly after production, the axions become non-relativistic and their dynamics can be described by the Schrödinger--Poisson system, the non-relativistic limit of the Klein--Gordon equation:
\begin{equation}\label{eq:schroedinger}
    \left(i\partial_t + \tfrac{3}{2}iH + \frac{\nabla^2}{2a^2m_a} - m_a(\Phi + \Phi_{\rm self})\right)\psi = 0 \, , 
    \qquad 
    \frac{\nabla^2 \Phi}{a^2} = 4\pi G \rho = 4\pi G (\rho_a + \rho_c + \rho_b)\,.
\end{equation}
Here the axion field $\phi$ is written in terms of the non-relativistic field $\psi$ as 
\begin{equation}
    \phi = \frac{1}{\sqrt{2m_a}}\psi e^{-im_a t} + {\rm c.c.}\,, \qquad \psi=\frac{\sqrt{\rho_a}}{m_a} e^{i\theta}\,,
\end{equation}
and the velocity field is $v_i = \partial_i\theta/(am_a)$.
$\Phi$ is the gravitational potential sourced by the total matter energy density $\rho=\rho_a+\rho_c+\rho_b$, $\rho_a = m_a |\psi|^2$ is the axion energy density, $\rho_c$ is the density of the additional non-interacting CDM component, and $\rho_b$ the baryon density. 
The term $\Phi_{\rm self}$ accounts for axion self-interactions.\footnote{For $\phi \ll f_a$ this is dominated by the quartic coupling contribution, $\Phi_{\rm self} = \lambda \rho_a / m_a^4$, with $\lambda \simeq -m_a^2/(8f_a^2)$.} 
During radiation and matter domination, including the transition between the two, the Hubble parameter is approximated by
\begin{equation}\label{eq:H2a}
    H^2(a) = \frac{H_{\rm eq}^2}{2} \left[ \left(\frac{a_{\rm eq}}{a}\right)^4 + \left(\frac{a_{\rm eq}}{a}\right)^3 \right] \, ,
\end{equation}
where quantities with subscript ``eq'' are defined at matter–radiation equality.

Eq.\,\eqref{eq:schroedinger} should be solved with initial conditions set by the axion field configuration sourced by strings/walls, with the power spectrum in Eq.\,\eqref{eq:Piso}. One might expect the resulting evolution to be similar to that of CDM. 
However, several effects, arising from the fact that axion particles exhibit wave-like behavior and are produced semi-relativistically, could significantly modify the dynamics between their production and the onset of clustering, during both matter- and radiation-dominated eras. 
These effects follow from Eq.\,\eqref{eq:schroedinger} and are discussed below for a generic axion fraction $f_{\rm DM}$, building on the recent work of Refs.~\cite{Amin:2025nxm,Amin:2025ayf}. Taking these into account is essential when assessing the impact of the enhanced small-scale structure from $\mathcal{P}_{\rm iso}$ on cosmological and astrophysical observables. As we will discuss next, for momenta larger than  (quantum/classical) Jeans scale, the axion isocurvature fluctuations entering Eq.~\eqref{eq:powerspec} do not grow; nevertheless, when \(f_{\rm DM}\!\ll\!1\) they continue to gravitationally source CDM perturbations, which then amplify and cluster—albeit with a progressively reduced amplitude at higher \(k\).

The total power spectrum of Eq. \eqref{eq:Piso}, which includes the axion and CDM component, evolves approximately as 
\begin{equation}\label{eq:Pak}
\mathcal{P}(a,k) = T_{\rm ad}^2(a,k)\,T_{\rm fs}^2(a,k)\,\mathcal{P}_{\rm ad}(k)+T_{\rm iso}^2(a) \mathcal{P}_{\rm iso}(k) \, .
\end{equation}
where $T_{\rm ad}(a,k)$ denotes the standard linear CDM adiabatic transfer function, $T_{\rm fs}$ the suppression of adiabatic perturbations from axion free streaming, sourced by the particles’ finite velocity dispersion,\footnote{See Refs.~\cite{Amin:2025sla,Amin:2025dtd,Amin:2022nlh,Amin:2025ayf,Amin:2025sla} for a more detailed derivation of Eq.~\eqref{eq:Pak}.} and $T_{\rm iso}$ the transfer function of the isocurvature component, encoding the Jeans-scale suppression, as discussed in the second part of this subsection.

\vspace{3mm}
\noindent
{\bf Free streaming.}    

\noindent
Free streaming is the process by which particles with a nonzero velocity dispersion move out of overdense regions into underdense ones. For axions  with comoving momentum $k_a$ (at the peak of the energy spectrum in Fig.~\ref{fig:Piso-sim} left), the typical velocity is $v=(k_a/a)/\sqrt{(k_a/a)^2+m_a^2}\simeq k_a/(a m_a)$. This leads to a comoving free-streaming length, defined as the distance traveled from the time $a_i$ when the particles become non-relativistic to some later time $a_f$:  
\begin{equation}\label{eq:kfs}
    k_{\rm fs}^{-1}(a_f)=\int_{t_i}^{t_f} \frac{dt}{a}\,v
    =\int_{a_i}^{a_f} \frac{da}{a^2 H}\,
    \frac{k_a/a}{\sqrt{(k_a/a)^2+m_a^2}}
    \;\simeq\; \frac{\sqrt{2}\,k_{a}}{m_a H_{\rm eq} a_{\rm eq}^2}\,
    \log\!\left(\frac{a_f}{a_i}\right)\, ,
\end{equation} 
where we used~Eq.\,\eqref{eq:H2a}. The last equality in Eq.\,\eqref{eq:kfs} holds for $t_f \lesssim t_{\rm eq}$; for $t_f > t_{\rm eq}$ the logarithm is replaced by $\log(4a_{\rm eq}/a_i)$. This shows that most of the free-streaming length is accumulated during radiation domination. Since $a_i=a_\star$ (more precisely, $a=a_\ell$), numerically the logarithm is $\log(a_{\rm eq}/a_\star) \simeq 5\!-\!11$ for $m_a=10^{-24}\!-\!10^{-18}\,\mathrm{eV}$. For postinflationary axions with $N=1$, 
we have $k_a\simeq k_{\rm wn}$, see  Fig.~\ref{fig:Piso-sim} and the discussion around it. Thus, from Eq.\,\eqref{eq:kfs}, one finds $k_{\rm fs}/k_\star \simeq [C\,\log(a_{\rm eq}/a_\star)]^{-1}$, up to relatively small corrections from changes in the number of relativistic degrees of freedom and the exact relation between $k_a$ and  $k_{\rm wn}$. Consequently, the free-streaming scale $k_{\rm fs}$ is always smaller than both $k_\star$ and the white-noise scale $k_{\rm wn}$.\footnote{With a slight abuse of notation, we will denote as $k_{\rm fs}$ the value of the function in Eq.~\eqref{eq:kfs} at $a = a_{\rm eq}$.}

Free streaming erases the initial correlations between momentum modes on length scales below the free-streaming scale, $k \gtrsim k_{\rm fs}$. 
For adiabatic perturbations, which are correlated across modes, this results in a suppression of power at $k \gtrsim k_{\rm fs}$. 
This is represented by the transfer function $T_{\rm fs}$ of Eq.\,\eqref{eq:Pak}, which is 
\begin{equation}\label{eq:T2ad}
T_{\rm fs}(a, k)= f_{\rm DM}\frac{\sin (k/k_{\rm fs}(a))}{k/k_{\rm fs}(a)}+(1-f_{\rm DM}) \,. 
\end{equation} 
 The first term corresponds to the suppression of adiabatic perturbations in the axion dark matter component, whereas the second simply indicates that the CDM adiabatic perturbations remain unaffected. Eq.\,\eqref{eq:T2ad} follows from a straightforward extension of the result in Ref.~\cite{Amin:2022nlh}, under the assumptions that (i) the additional CDM component is uncorrelated with respect to the axion one and carries only standard adiabatic perturbations, and (ii) the axion spectrum is a delta function peaked at $k_{\rm wn}$, thereby simplifying the integral that enters $T_{\rm fs}$ in Ref.~\cite{Amin:2022nlh}.\footnote{The result in Eq.\,\eqref{eq:T2ad} is consistent with the recent derivation presented in Refs.~\cite{Amin:2025sla,Amin:2025ayf}, which instead 
 solves the equations for the particle-distribution functions.} This second assumption leads to unphysical oscillations of $T_{\rm fs}$ in $k$, related to the $\sin(k/k_{\rm fs})$ factor, which disappear when a smoother axion spectrum is considered.

\begin{figure}[t]
    \centering
        \includegraphics[width=0.85\textwidth]{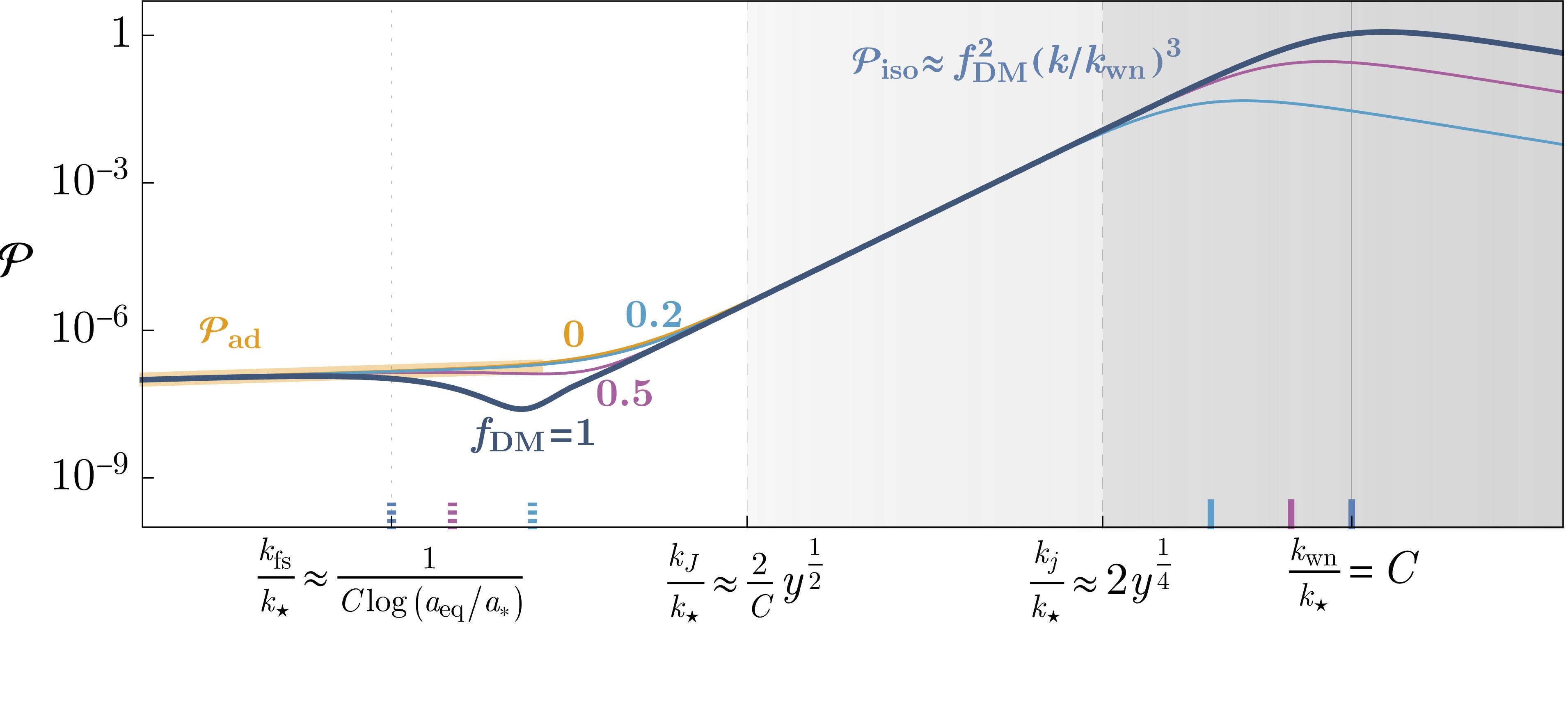} \vspace{-4mm}
     \caption{{\small A sketch of the dimensionless power spectrum $\mathcal{P}=\mathcal{P}_{\rm ad}+\mathcal{P}_{\rm iso}$ around matter-radiation equality, shown as a function of comoving momentum $k$ in units of $k_\star=m_a a_\star\simeq 54 (m_a/10^{-20}\,\mathrm{eV})^{1/2}\,\mathrm{Mpc}^{-1}$. 
     The isocurvature contribution $\mathcal{P}_{\rm iso}$ peaks at $k_{\rm wn}=Ck_\star$ with $C=\mathcal{O}(10)$. 
     For $f_{\rm DM}=1$, axion free streaming suppresses the adiabatic component $\mathcal{P}_{\rm ad}$ at $k \gtrsim k_{\rm fs}$, visible as the dip in the dark blue curve, while leaving the white-noise tail of $\mathcal{P}_{\rm iso}$ unaffected. This suppression becomes increasingly smaller for subdominant axion fractions $f_{\rm DM}=0,0.1,0.5$ (orange, light blue, and purple curves, respectively). 
     In addition, during the later evolution at $a\gtrsim a_{\rm eq}$ (not shown),} the axion velocity dispersion (classical pressure) and quantum pressure inhibit the growth of the axion fluctuations at $k \gtrsim k_J \propto (a/a_{\rm eq})^{1/2}\equiv y^{1/2}$ and $k \gtrsim k_j \propto y^{1/4}$ (light and dark gray regions).   
     }
     \label{fig:Psketch}
\end{figure}

The impact of free streaming on the adiabatic power spectrum is illustrated in Fig.~\ref{fig:Psketch}, where we show the total matter spectrum as well as its individual components for different choices of $f_{\rm DM}$. For visualization purposes in plotting $\mathcal{P}$ 
we keep the amplitude of $\mathcal{P}_{\rm iso}\propto f_{\rm DM}^2 (k/k_{\rm wn})^3$ fixed by simultaneously decreasing $f_{\rm DM}$ and $k_{\rm wn}$, so that the peak of $\mathcal{P}$, marked by solid ticks, appears at progressively smaller amplitudes. If $f_{\rm DM}=1$, Eq.\,\eqref{eq:T2ad} predicts a strong suppression of adiabatic modes at $k\gtrsim k_{\rm fs}(a_{\rm eq})$, visible as a dip (blue curve) compared to the CDM case without free streaming (orange curve).
In contrast, for $f_{\rm DM}\ll1$, the effect of free-streaming is present only in the axion component and becomes negligible. Because $k_{\rm wn}$ is also reduced, the free-streaming scale (dashed ticks) shifts to increasingly UV scales as $f_{\rm DM}$ decreases, see Eq.~\eqref{eq:kfs}.

\begin{figure}[t]
    \centering
    \includegraphics[width=0.42\textwidth]{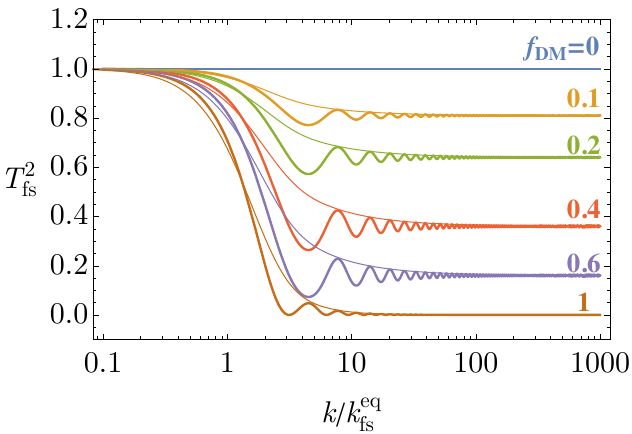}
   \ \ 
        \includegraphics[width=0.485\textwidth]{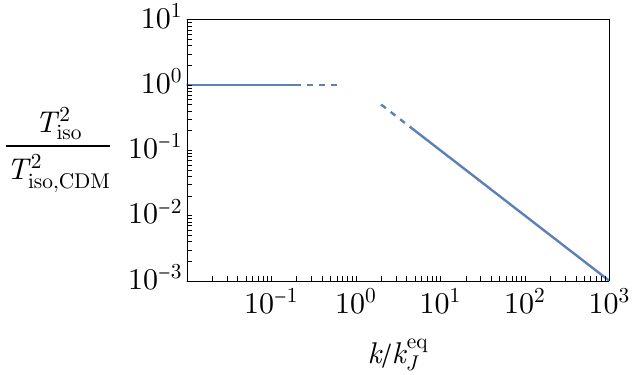}
     \caption{{\small {\bf{Left}}: 
     The free-streaming transfer function $T^2_{\rm fs}$ for the adiabatic power spectrum $\mathcal{P}_{\rm ad}$ for different values of the fraction $f_{\rm DM}$ of dark matter in axions, for a delta function axion spectrum (thick) and a more realistic one (thin envelope).  For $f_{\rm DM}\ll1$, only the adiabatic perturbations in the (warm) axion dark matter component are affected by free streaming and $T^2_{\rm fs}\simeq 1-2f_{\rm DM}$ for $k\gtrsim k_{\rm fs}(a_{\rm eq})$;  free-streaming effect is increasingly negligible.  {\bf{Right}}: Sketch of the isocurvature transfer function for $f_{\rm DM}\ll1$ and at $a\gg a_{\rm eq}$, normalized to that expected in pure CDM. For $k\gtrsim k_J^{\rm eq}$ the axion velocity dispersion mildly suppresses the growth of the axion+CDM fluctuations, leading to $T_{\rm iso}^2\propto 1/k$.} 
      \label{fig:k12} }
    
\end{figure}

For completeness, Fig.~\ref{fig:k12} (left) shows the suppression factor $T_{\rm fs}^2$ for different values of $f_{\rm DM}$. The thick lines correspond to $T_{\rm fs}$ in Eq.~\eqref{eq:T2ad}, which exhibits unphysical oscillations, while the thin lines show the same transfer function (Eq.~(5) of Ref.~\cite{Amin:2022nlh}) evaluated for a (smoother) spectrum resembling Fig.~\ref{fig:Piso-sim} (left). Additionally, it is worth noting that if strings and domain walls generate correlations between momentum modes near the spectral peak or at higher momenta, they would also be erased by free streaming, resulting in an additional suppression factor in the last term of Eq.\,\eqref{eq:Pak} for the relevant $k$. The white-noise component, in contrast, is unaffected, so $\mathcal{P}_{\rm iso}$ remains peaked at the original momentum scale.

\vspace{3mm}
\noindent
{\bf Jeans suppression of perturbation growth.} \nopagebreak  

\noindent
As the Universe enters matter domination, axion dark matter self-gravity, i.e. the $-m_a\Phi\psi$ term in Eq.\,\eqref{eq:schroedinger}, becomes important. 
The $\mathcal{O}(1)$ fluctuations at the peak scale $k_{\rm wn}$ are expected to collapse into compact objects known as \emph{axion miniclusters}~\cite{Hogan:1988mp,Kolb:1993zz,Zurek:2006sy,Hardy:2016mns,OHare:2021zrq,Chang:2024fol}, with characteristic mass $\sim (\pi/k_{\rm wn}^3)\,\rho_a(a_{\rm eq})$, while modes with $k<k_{\rm wn}$, initially perturbative, grow linearly like CDM before eventually clumping into larger minihalos. 
However, this gravitational collapse and growth are suppressed for fluctuations with wavelengths below the relevant Jeans scales, due to the effects of the classical velocity dispersion and their quantum pressure.  
To obtain the parametric dependence of these scales,  we rewrite Eq.\,\eqref{eq:schroedinger} in terms of the axion density  $\rho_a$ and velocity $v_i$  fields as:  
\begin{equation}\label{eq:continuityEuler}
    \dot{\rho}_a + 3H\rho_a + \frac{\partial_i(\rho_a v_i)}{a} = 0 \, , 
    \quad \dot{v}_i + H v_i + \frac{v_j \partial_j v_i}{a} = - \frac{\partial_i(\Phi + \Phi_Q + \Phi_{\rm self})}{a} \, , 
    \quad \frac{\nabla^2\Phi}{a^{2}} = 4\pi G \big( \rho_a + \rho_{c} + \rho_b \big) \, .
\end{equation}
Evidently, the axion dark matter dynamics follow the standard continuity, Euler, and Poisson equations, with the addition of a quantum pressure potential,  
\begin{equation}
    \Phi_Q \equiv - \frac{\nabla^2\sqrt{\rho_a}}{2m_a^2 a^2 \sqrt{\rho_a}}\,,
\end{equation}
which arises from the wave nature of the axion field on small spatial scales. 
In the following we neglect $\Phi_{\rm self}$, which in our case is only relevant during radiation epoch~\cite{Gorghetto:2022ikz}.\footnote{Self-interactions can be important in other regimes, such as in the case of sizable attractive couplings—where they modify core–halo structure, affect soliton masses and density profiles, and can even trigger collapse above a critical threshold; see, e.g., Ref.~\cite{Painter:2024rnc}.} Note that, the fluid description becomes singular at points where $\rho_a=0$ (as in a random superposition of waves), so we use it only in a coarse-grained sense: for perturbations around a homogeneous background and away from such nodes.

The Jeans suppression can be illustrated by tracking the evolution of perturbations $\rho_a$ and $\rho_c$, defined in Eq.\,\eqref{eq:delta}, in the axion and additional CDM component, which follows Eq.\,\eqref{eq:continuityEuler} with $\Phi_Q=\Phi_{\rm self}=0$. 
Expanding Eq.\,\eqref{eq:continuityEuler} to linear order in $\delta_a$, $\delta_c$, and $v_i$ yields the evolution equations:  
\begin{align}\label{eq:delta1}
    \ddot{\delta}_a + 2H\dot{\delta}_a + c_s^2 \frac{k^2}{a^2}\,\delta_a - 4\pi G \bar{\rho}\,\delta &= 0 \, , 
    & c_s^2 = \frac{k^2}{4m_a^2 a^2} + \frac{k_a^2}{a^2} \, , \\ \label{eq:delta2}
    \ddot{\delta}_c + 2H\dot{\delta}_c - 4\pi G \bar{\rho}\,\delta &= 0 \, ,
\end{align}
where $\delta_a$ and $\delta_c$ are expressed in Fourier space. Relative to CDM, axion perturbations acquire an effective sound speed $c_s^2$ with two contributions: the first arises from quantum pressure, while the second from the axion velocity dispersion. 
The latter can be derived by solving the Schrödinger–Poisson system for a superposition of modes with momentum spread $k_a^2/a^2\simeq k_{\rm wn}^2/a^2$~\cite{Amin:2025sla}.

If $f_{\rm DM}=1$, Eq.\,\eqref{eq:delta1} can be simplified by introducing $y=a/a_{\rm eq}=(1+z_{\rm eq})/(1+z)$. Neglecting baryons, $\rho_b=0$, and defining $\delta'=d\delta/dy$, we get
\begin{equation}\label{eq:delta_3}
\delta''+\frac{3 y+2}{2 y (y+1)}\delta'
-\frac{3}{2 y (y+1)}
\left[ 1-\left(\frac{k}{k_{j}^{\rm eq}}\right)^4\frac{1}{y}
-\left(\frac{k}{k_{J}^{\rm eq}}\right)^2\frac{1}{y}\right]\delta = 0 \, ,
\end{equation}
where the comoving quantum and classical Jeans scales are defined by  
\begin{equation}\label{eq:kJs}
k_j \equiv a \left(16\pi G  \bar{\rho} m_a^2 \right)^{1/4} \propto a^{1/4} \,, 
\qquad 
k_J \equiv a \frac{(4\pi G \bar{\rho} m_a^2)^{1/2}}{k_a/a} \propto a^{1/2} \, .
\end{equation}
From Eq.\,\eqref{eq:delta_3}, for scales larger than the Jeans lengths, $k\ll k_j, k_J$, the additional pressure terms are negligible and perturbations grow as in CDM, $\delta\propto 1+(3/2)\,(a/a_{\rm eq})$. This growth is encoded in the isocurvature transfer function 
\begin{equation}\label{eq:Tiso}
T_{\rm iso} (a) = \left(1+\frac{3\gamma}{2\alpha_-} \frac{a}{a_{\rm eq}}\right)^{\alpha_-}\, \qquad \quad \text{for} \ \ \ \ k\lesssim k_J^{\rm eq} \ \ , 
\end{equation}
where we allowed $\rho_b\neq0$ and $\alpha_- = {\frac{\sqrt{1+24\gamma}-1}{4}}$ and 
$\gamma = (\bar{\rho}_c+\bar{\rho}_a)/\bar{\rho} \sim 0.265/0.315$ is the fraction of matter that can collapse gravitationally. However, for $k$ above the Jeans scales, quantum and classical pressure lead instead to oscillatory solutions, and prevent mode growth. 

Crucially, at matter-radiation equality the classical and quantum Jeans momenta are parametrically close to the typical emission scale $k_{\rm wn}=Ck_\star$ where $\mathcal{P}_{\rm iso}$ peaks. Indeed,
\begin{equation}\label{eq:KJkj}
 \frac{k_j}{k_\star}=\frac{k_{j}^{\rm eq}}{k_\star}\left(\frac{a}{a_{\rm eq}}\right)^{1/4}\simeq 2\left(\frac{a}{a_{\rm eq}}\right)^{1/4} , 
 \qquad   
 \frac{k_J}{k_\star}=\frac{k_{J}^{\rm eq}}{k_\star}\left(\frac{a}{a_{\rm eq}}\right)^{1/2}\simeq \frac{2}{C}\left(\frac{a}{a_{\rm eq}}\right)^{1/2} \, ,
\end{equation}
where we used $k_j^{\rm eq}/k_\star \sim 2$ and $k_j^2 = 2k_a k_J$ following from Eq.\,\eqref{eq:kJs}, $a^2_\star/a_{\rm eq}^2\simeq H_{\rm eq}/H_\star$ and $k_a\simeq k_{\rm wn}$.
We display the scales $k_j$ and $k_J$ in Fig.~\ref{fig:Psketch}. At this time the classical Jeans momentum $k_J$ is smaller than the quantum one by a factor $\sim 1/C$, so the dominant suppression comes from the velocity dispersion. For $C=\mathcal{O}(10)$, the modes with $k\gtrsim k_J$ lie well below the peak $k_{\rm wn}$ and are unable to grow (gray region in Fig.~\ref{fig:Psketch}). This includes the large density fluctuations at $k_{\rm wn}$, implying that the power spectrum at lowest modes that can collapse is $\mathcal{P}_{\rm iso}({k=k_{J}})
\sim (C^{-2})^3\sim 10^{-6}$ and axion miniclusters do not form.\footnote{In contrast, if $C=\mathcal{O}(1)$, then $k_{\rm wn}\simeq k_j^{\rm eq}\simeq k_J^{\rm eq}$, and such fluctuations would indeed collapse into miniclusters (or, more precisely, axion stars)~\cite{Gorghetto:2022sue,Gorghetto:2024vnp}. The classical Jeans scale $k_J$ overtakes the quantum scale $k_j$ only at very late times, $a/a_{\rm eq} \sim (2C)^4 = \mathcal{O}(10^4)$, i.e. well after the present epoch. Hence, throughout the cosmologically relevant period, classical pressure dominates over quantum pressure.}
We do not further discuss the growth of isocurvature fluctuations at $k > k_J^{\rm eq}$ for the $f_{\rm DM} = 1$ case, since for values of $m_a$ not excluded by free streaming (see Sec.\,\ref{sec:results}), the resulting enhancement of structure formation occurs at amplitudes or spatial scales too small to be probed by LSS data. In other words, free streaming likely provides the dominant constraint in the case of axions constituting the primary dark matter component.

The situation changes if the axion is a subdominant dark matter component,  $f_{\rm DM}\ll1$. In this case, an axion isocurvature perturbation at $k>k_J$ sources an isocurvature CDM perturbation at that mode, $(1 - f_{\rm DM})\delta_c\simeq \delta_c$.\footnote{$k_J$ is defined as in Eq.~\eqref{eq:KJkj}, with $\bar{\rho}$ corresponding to the total dark matter density (rather than only the subdominant component $\bar{\rho}_a$). Intuitively, this is because $k_J$ is obtained by equating the interaction timescale $k^{-1}/c_s$ with the free-fall time $1/\sqrt{G\bar{\rho}}$, where $\bar{\rho}$ accounts for all matter contributing to the gravitational potential.}  Once this grows to the same order as the axion perturbation $f_{\rm DM}\delta_a$, the influence of the axion becomes negligible, and the dominant component follows 
\begin{equation}\label{eq:delta_4}
\delta_c''+\frac{3 y+2}{2 y (y+1)}\delta_c'-\frac{3(1-f_{\rm DM})}{2 y (y+1)}\delta_c =0 \, ,
\end{equation}
undergoing CDM-like growth for that mode. As shown in Refs.~\cite{Amin:2025nxm,Amin:2025ayf}, $\delta_c$ does not become comparable to $f_{\rm DM}\delta_a$ immediately at matter-radiation equality, but at a later time when the scale factor is increased by  $\sqrt{k / k_J^{\rm eq}}$. This result can be rigorously derived by following the phase-space distribution function of the particles in the mixed warm and cold dark matter system. The obtained transfer function reads
\begin{equation}\label{eq:TisoJ}
T_{\rm iso} (a,k) \simeq 1+\frac{y^2}{k/k_J^{\rm eq}}\qquad \text{for} \quad k\gtrsim k_J^{\rm eq} \,  .
\end{equation}
The above equation can be understood as follows: at the dynamical time $t_D \simeq H^{-1} \simeq (G\bar{\rho})^{-1/2}$, gravity effectively probes a field with an amplitude of order  $f_{\rm DM}\,\delta_a \big/ (\Delta t / t_D)^{1/2}$,
where $\Delta t \sim k^{-1}/(k_a/m)$ is the timescale over which the average dark-matter field changes by unity within a region of size $1/k$.  
The factor $(\Delta t / t_D)^{1/2}$ accounts for the number of statistically independent field oscillations that occur during one dynamical time, thus reducing the effective amplitude seen by gravity over $t_D$. As a result, the isocurvature transfer function scales as $T_{\rm iso} \propto 1/k$ for $k \gtrsim k_J^{\rm eq}$—a much milder suppression than the $1/k^4$ behavior occurring for purely (warm) axion dark matter~\cite{Amin:2025dtd}. We also note that the presence of the quantum Jeans scale $k_j$ does not alter the above scaling in Eq.~\eqref{eq:TisoJ}~\cite{Amin:2025nxm}. As we will discuss in the next section, the different scaling of the transfer function in Eq.\,\eqref{eq:TisoJ} with respect to CDM renders standard analyses and bounds not directly applicable. We will outline a conservative way to address this.

\section{Large-scale structure probes
}
\label{sec:LSS}

\begin{figure}
    \centering
    \includegraphics[width=0.65\linewidth]{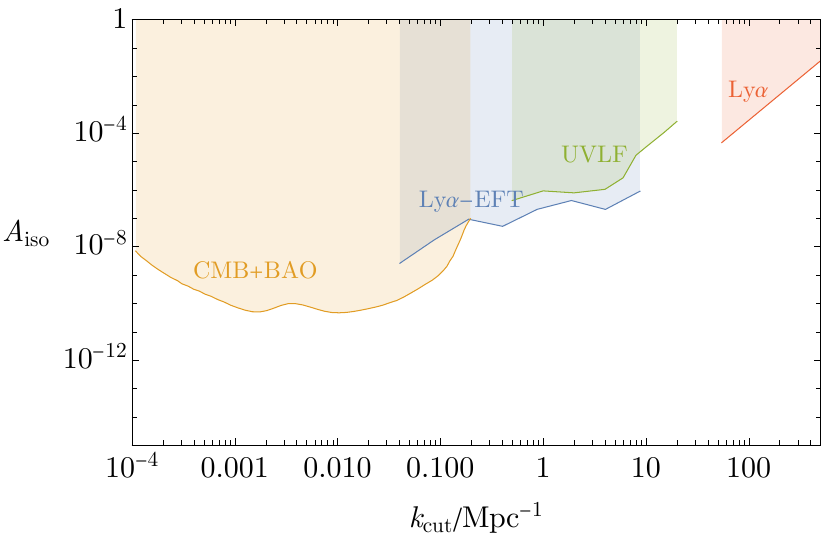}
    \caption{95\% C.L. exclusion limits on the primordial amplitude $A_{\rm iso}$ of the isocurvature component of the power spectrum $\mathcal{P}_{\rm iso}(k)=A_{\rm iso}(k/k_{\rm cut})^3$ as a function of its UV cutoff $k_{\rm cut}$. 
    The bounds originate from CMB+BAO measurements~\cite{Buckley:2025zgh} (orange),  Lyman-$\alpha$ MIKE/HIRES data with EFT-based~\cite{Ivanov:2025pbu} (blue) and hydrodynamic simulations~\cite{Murgia:2019duy} (red), as well as the UVLF of HST galaxies (green) used in this work. The bounds assume all dark matter fluctuations evolve as CDM, i.e. with negligible velocity dispersion and quantum pressure. }
    \label{fig:Piso}
\end{figure}

\begin{figure}
    \centering
    \includegraphics[width=0.65\linewidth]{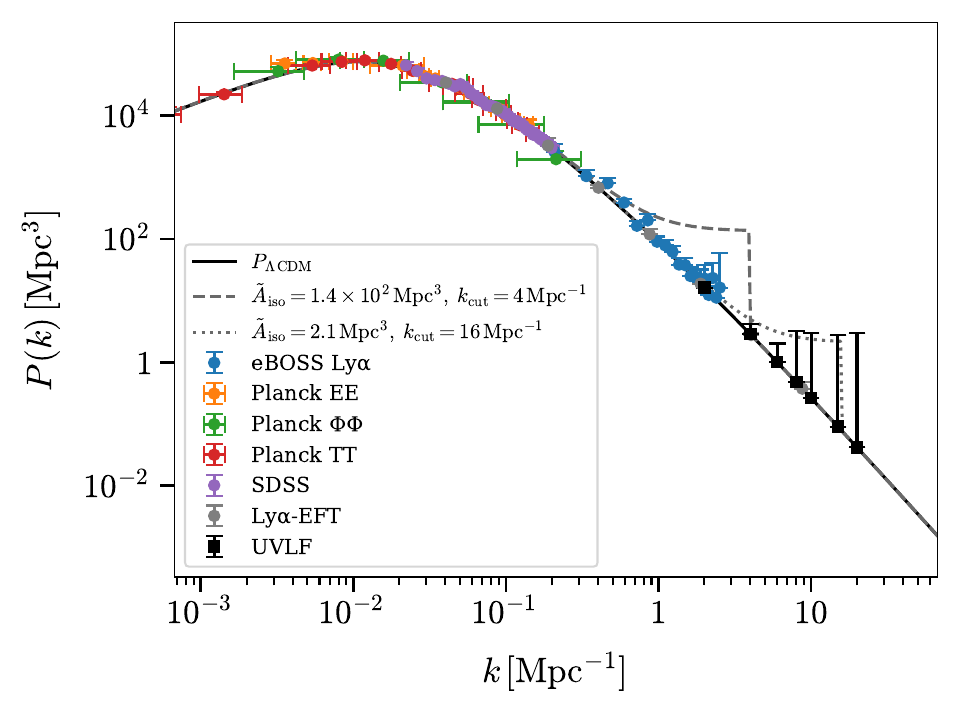}
    \caption{\small{Linear matter power spectrum $P(k)=(2\pi^2/k^3)\mathcal{P}(k)$ at $z=0$ as a function of the comoving momentum $k$. The $\Lambda$CDM prediction (black) is compared to two benchmark models with additional isocurvature white-noise components (gray, dashed and dotted). Black points with error bars show the 95\% C.L. constraints on excess power inferred in this work from Hubble Space Telescope UVLF data. Blue points denote Lyman-$\alpha$ measurements from eBOSS DR14~\cite{eBOSS:2018qyj,Chabanier:2019eai}, and gray points the Lyman-$\alpha$ EFT analysis~\cite{Ivanov:2025pbu} (which constrains an isocurvature enhancement as well). The remaining error bars correspond to SDSS DR7 luminous red galaxies (purple)~\cite{2010MNRAS.404...60R} and the \textit{Planck} 2018 CMB power spectra—temperature (red), polarization (orange), and lensing (green)~\cite{Planck:2018vyg}.}}
    \label{fig:Ptot}
\end{figure}

The enhancement of the matter power spectrum in the axion post-inflationary  scenario gives rise to a non-standard structure formation, leading to an increased abundance of dark matter halos capable of hosting galaxies and early stars~\cite{Irsic:2019iff}. In this section, we first present our novel constraint derived from this process, based on the analysis of the UVLF dataset obtained by the HST over the redshift range $4-10$~\cite{Bouwens:2014fua,2015ApJ...810...71F,Atek:2015axa,Livermore:2016mbs,2017ApJ...843..129B,2017ApJ...838...29M,2018ApJ...854...73I,2018ApJ...855..105O,Atek:2018nsc,2020ApJ...891..146R,2021AJ....162...47B}. 
We then collect all other leading constraints from LSS observations. The outcome of this analysis is shown in Fig. \ref{fig:Piso} as exclusion limits on the dimensionless coefficient $A_{\rm iso}$ of a generic primordial  white-noise isocurvature  power spectrum with cutoff momentum $k_{\rm cut}$:
\begin{equation}\label{eq:Piso1}
    \mathcal{P}_{\rm iso}(k)=A_{\rm iso}(k/k_{\rm cut})^3\equiv \tilde{A}_{\rm iso}k^3/(2\pi^2) \,,
\end{equation}
where we also defined the dimensionful offset $\tilde{A}_{\rm iso}$ for practical purposes. The corresponding datapoints with errorbars on the total dimensionful linear power spectrum $P(k)$ today are shown in Fig.~\ref{fig:Ptot}.

\vspace{3mm}
\noindent
{\bf Subtleties in the application to the axion.} \nopagebreak  

\noindent
The limits on $\mathcal{P}_{\rm iso}$ presented in this section assume standard CDM growth. However, as discussed in Sec.~\ref{sec:scales} and shown in Fig.~\ref{fig:k12} (right), the Jeans suppression reduces the growth of axion dark matter perturbations for $k > k_J^{\rm eq}$, including near the peak at $k \simeq k_{\rm wn}$. As a result, the bounds cannot be directly applied to the axion with the naive identification $A_{\rm iso}=f_{\rm DM}^2$ and $k_{\rm cut}=k_{\rm wn}$, which would follow from comparing Eqs.\,\eqref{eq:Piso} and~\eqref{eq:Piso1}. 

Still, axion isocurvature fluctuations behave as CDM for $k \lesssim k_J^{\rm eq}$. A conservative lower bound on the resulting enhanced structures can therefore be obtained by considering the  spectrum only for $k \lesssim k_J^{\rm eq}$. This corresponds to assuming $k_{\rm cut}=k_J^{\rm eq}$. 
However, when $f_{\rm DM}\ll 1$, this choice becomes overly restrictive because it discards most of the white-noise power (at $k>k_J^{\rm eq}$) that still seeds structures on the CDM component; its effect is suppressed only by a factor $\sim k_J^{\rm eq}/k$,  see Eq.\,\eqref{eq:TisoJ}. We account for this by taking $k_{\rm cut}=k_{\rm wn}$ and multiplying the isocurvature amplitude $A_{\rm iso}=f_{\rm DM}^2$ by $k_J^{\rm eq}/k_{\rm wn}\simeq 2C^{-2}$. This is still a conservative choice, since only modes near $k_{\rm wn}$ experience this level of suppression, while modes at lower $k$ are less suppressed.

In the remainder of this section we treat $A_{\rm iso}$ and $k_{\rm cut}$ as free independent parameters. In Sec.~\ref{sec:results} we apply the above strategy to constrain axion mass and decay constant.

\subsection{Halo Mass Function}
\label{sec:HMF}

The halo mass function (HMF) quantifies the comoving number density of dark matter halos $n_h$ as a function of their mass $M_h$ and redshift $z=1/a-1$. It provides the fundamental link between the matter power spectrum and the abundance of galaxies, and is key in our modeling of LSS. The goal of this section is to calculate the HMF for a model that includes the white-noise contribution in Eq.\,\eqref{eq:Piso1} on top of the adiabatic part. While the HMF and galaxy abundances should be, in principle, determined through cosmological simulations, we approximate them using well-established semi-analytical methods, as detailed below. 

We define the effective radius $R$ enclosing a halo of mass $M_h$ at redshift $z$, assuming a mean dark matter density $\bar{\rho}$:
\begin{equation}\label{eq:radiusdef}
R = \left(\frac{3 M_h}{4 \pi \bar{\rho}} \right)^{\!1/3}\,.
\end{equation}
We work with the dimensionful matter power spectrum at redshift $z$,
\begin{equation}
P(k,z) \equiv \frac{2\pi^2}{k^3}\,\mathcal P(k,z)\,,
\end{equation}
where $\mathcal P(k,z)$ is the dimensionless power spectrum defined in Eqs.~\eqref{eq:powerspec} and evolving in the linear regime as in Eq.\,\eqref{eq:Pak}. $\mathcal P(k,z)$ contains the $\Lambda$CDM contribution ($\mathcal{P}_{\rm ad}$) and the isocurvature one ($\mathcal{P}_{\rm iso}$). To evaluate the linear evolution of the former, we employ the Boltzmann solver  \textsc{CLASS}~\citep{2011arXiv1104.2932L} to compute $T_{\rm ad}$, as detailed in Sec.\,\ref{sec:UVLF}, while we neglect $T_{\rm fs}$ in Eq.\,\eqref{eq:Pak}, which is a good approximation for $f_{\rm DM} \ll 1$. 
Similarly, the isocurvature part is evolved as in Eqs.\,\eqref{eq:Pak} and \eqref{eq:Tiso} for all $k$, i.e. neglecting the effect of velocity dispersion of Eq.\,\eqref{eq:TisoJ}. 

The variance of the density fluctuation on scale $R$ and redshift $z$ is then
\begin{equation}\label{eq:variance}
\sigma^2(M_h,z) = \int \frac{dk\,k^2}{2\pi^2}\, W^2(kR)\, P(k,z)\, ,  
\end{equation}
where $W$ is a window function. The commonly-adopted top-hat filter form
\begin{equation}\label{eq:tophat}
W(kR) = \frac{3\left[\sin(kR)-kR\cos(kR)\right]}{(kR)^3}\,,
\end{equation}
is known to overestimate the abundance of halos for power spectra with a high-$k$ cutoff \citep{2013MNRAS.428.1774B,Schneider:2014rda}, as is the case with our axion models. To avoid this issue, we follow \citep{Bertschinger:2006nq,Schneider:2013ria,Sabti:2021unj} in adopting a sharp-$k$ window function instead
\begin{equation}\label{eq:sharpk}
W(kR) = \Theta\left( 1-k R\right)\,,
\end{equation}
where $\Theta$ denotes the Heaviside step function. Furthermore, to ensure a well-defined relationship between halo mass and radius when using $W$ in Eq.\,\eqref{eq:sharpk}, we introduce an additional scaling parameter, $c_{\rm CM}$, which redefines the radius from Eq.~\eqref{eq:radiusdef} to $c_{\rm CM} R$ \citep{Schneider:2014rda,Sabti:2021unj}. We adopt the value $c_{\rm CM}=2.5$ which has been found to provide a good fit to a wide range of dark matter simulations \citep{Schneider:2014rda}.

\begin{figure}[htbp]
    \centering
    \includegraphics[width=0.5\linewidth]{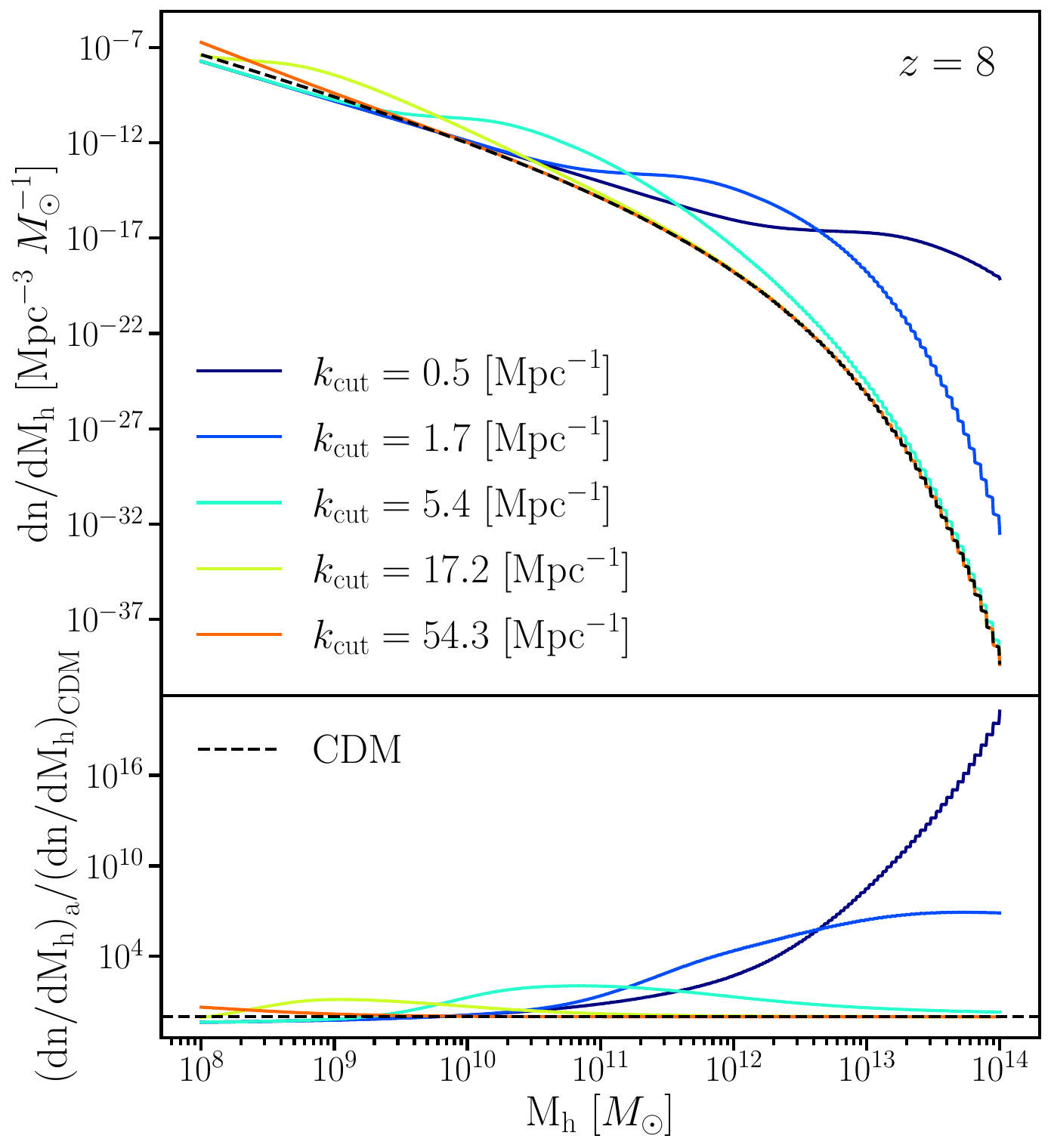} 
    \caption{{\small Halo mass function for standard CDM (black dashed) and for isocurvature enhancements corresponding to varying values of characteristic cutoff wavenumber, $k_{\rm cut}$, at $z=8$. The values $k_{\rm cut}=\{0.5,1.7,5.4,17.2,54.3\}$ $\rm Mpc^{-1}$ correspond to $\tilde{A}_{\rm iso}=\{1.2\cdot10^{-2},3.9\cdot10^{-4},1.2\cdot10^{-5},3.9\cdot10^{-7},1.2\cdot10^{-8}\}$ $\rm Mpc^{3}$, respectively. The lower panels show the fractional difference with respect to the standard CDM prediction.}}  \label{fig:halmasscomparison} 
\end{figure}

According to Press–Schechter theory and its extensions \cite{1974ApJ...187..425P,1991ApJ...379..440B}, halos correspond to regions where the smoothed density field exceeds a critical threshold for collapse. For an Einstein–de Sitter Universe this value is $\delta_{\rm cr}=1.686$ \cite{1986ApJ...304...15B,1991ApJ...379..440B}, which remains a good approximation for $\Lambda$CDM. A convenient way to characterize the likelihood of collapse on a given mass scale is through the ratio of the critical overdensity to the typical size of fluctuations. This dimensionless quantity, known as the peak significance, encodes how \emph{rare} a fluctuation of mass $M_h$ must be in order to collapse at redshift $z$:
\begin{equation}\label{eq:peak}
\nu_c(M_h,z) \equiv \frac{\delta_{\rm cr}}{\sigma(M_h,z)}\,.
\end{equation}
With these ingredients, the HMF is
\begin{equation}\label{eq:HMF}
\frac{dn_h}{d M_h} = \frac{\bar{\rho}}{M_h^2}\,
\nu_c(M_h)\, f\!\left(\nu_c(M_h)\right)\,
\frac{d\ln \nu_c(M_h)}{dM_h}\,,
\end{equation}
where $f(\nu)$ is the multiplicity function encoding the fraction of mass collapsed into halos. Importantly, all dependence on new physics enters through the matter power spectrum $P(k)$, and hence through $\nu_c$ as defined in Eqs.\,\eqref{eq:peak} and \eqref{eq:variance}. The (axion-induced) white-noise contribution at smaller scales, discussed in Sec.~\ref{sec:scales}, only modifies the HMF expression via $P(k)$.   
In the original Press–Schechter theory \cite{1974ApJ...187..425P}, this function follows a Gaussian distribution
\begin{equation}\label{eq:psmult}
f_{\rm PS}(\nu) = \sqrt{\frac{2}{\pi}}\,e^{-\nu^2/2},
\end{equation}
which is exact for an Einstein–de Sitter cosmology with a scale-free power spectrum, but fails to match $N$-body results at the accuracy required today. A more precise description is given by the Sheth–Tormen form \cite{Sheth:1999mn,Sheth:1999su}
\begin{equation}\label{stmult}
 f_{\rm ST}(\nu) = \sqrt{\frac{2}{\pi}}\,A(p)
\left[1+\frac{1}{(q\nu^2)^p}\right]\sqrt{q}\,
e^{-q\nu^2/2}\,,
\end{equation}
with normalization
\begin{equation}
A(p) = \Bigg[1+\pi^{-1/2}2^{-p}\Gamma\!\left(\tfrac{1}{2}-p\right)\Bigg]^{-1}\,,
\end{equation}
and free parameters $q,p$ fitted to simulations. The standard choices are $(q,p)=(0.707,0.3)$ or $(0.75,0.3)$, while for high-redshift applications $(q,p)=(0.85,0.3)$ provides a better fit \cite{Schneider:2020xmf}. Given, however, that in our chosen implementation the halo radius was previously redefined through the scaling parameter $c_{\rm CM}=2.5$, which has already been fitted to simulations, we set $q=1.0$ here to avoid redundancy, and work with $(q,p)=(1.0,0.3)$ as in Ref.~\citep{Sabti:2021unj}.\footnote{Alternative calibrations of the HMF, have also been proposed~\cite{Jenkins:2000bv, Reed:2006rw, Tinker:2008ff}, but they induce only minor variations in UVLF predictions \cite{Sabti:2021unj,Winch:2024mrt}. In this work, we therefore adopt the Sheth–Tormen form, consistent with the default GALLUMI implementation.} In Fig.~\ref{fig:halmasscomparison}, we illustrate the dependence of the halo mass function on the cutoff wavenumber $k_{\rm cut}$.  
As expected from Eq.~\eqref{eq:Piso1},  for fixed $A_{\rm iso}$, larger $k_{\rm cut}$ values lead to decreasing deviations (w.r.t. CDM), which in turn peak at increasingly smaller scales (and halo masses).

\subsection{Ultraviolet luminosity function}
\label{sec:UVLF}

The UVLF provides a statistical description of the abundance of galaxies as a function of their rest-frame UV magnitude. It is defined as the comoving number density of galaxies per unit magnitude, 
\begin{equation}
    \Phi_\mathrm{UV}(M_\mathrm{UV},z) \equiv \frac{d n_\mathrm{gal}}{d M_\mathrm{UV}}\, ,
\end{equation}
where $M_\mathrm{UV}$ is the absolute UV magnitude (typically measured at wavelengths around $1500\,\text{\AA}$), and $n_\mathrm{gal}$ denotes the cumulative number density of galaxies brighter than a given magnitude.  
The UVLF is one of the most important probes of early galaxy formation, as it directly encodes the efficiency of star formation in dark matter halos. In this section we outline the astrophysical model we use to predict it based on Refs.~\citep{Sabti:2021unj,Sabti:2021xvh}, and apply it to the isocurvature power spectrum.

Since galaxies are hosted in dark matter halos, the UVLF can be constructed from the HMF $dn_h/dM_h$ in Eq. \eqref{eq:HMF} and the conditional probability $\Pi(M_\mathrm{UV}|M_h,z)$ that a halo of mass $M_h$ hosts a galaxy with UV magnitude $M_\mathrm{UV}$:
\begin{equation}
    \Phi_\mathrm{UV}(M_\mathrm{UV},z) 
    = \int dM_h \, \frac{dn_h}{dM_h}(M_h,z) \, \Pi(M_\mathrm{UV} \,|\, M_h,z)\,.
    \label{eq:UVLF_construct_general}
\end{equation}
The conditional probability is modeled as a Gaussian distribution~\cite{Yang:2002ww,Sabti:2021xvh},
\begin{equation}
     \Pi(M_\mathrm{UV}|M_h,z) = 
    \frac{1}{\sqrt{2\pi}\,\sigma_{M_\mathrm{UV}}}
    \exp\!\left[-\frac{\big(M_\mathrm{UV}-\bar{M}_{\mathrm{UV}}(M_h,z)\big)^2}{2\sigma_{M_\mathrm{UV}}^2}\right]\,,
    \label{eq:UVprob_gaussian}
\end{equation}
where $\bar{M}_{\mathrm{UV}}(M_h,z)$ is the mean UV magnitude predicted by the galaxy-halo connection described below, and $\sigma_{M_\mathrm{UV}}$ quantifies the intrinsic statistical scatter in this relation due to processes such as bursty star formation, and is treated as a nuisance parameter that is fit to the data.

\vspace{0.5em}
\noindent
\textbf{The galaxy-halo connection.}  

\noindent
To evaluate the mean relation $\bar{M}_{\mathrm{UV}}(M_h,z)$, we construct a chain of mappings that connects halo mass to UV luminosity of galaxies hosted by it.   The steps are as follows:

\begin{enumerate}
    \item   
    The mean UV magnitude is related to the mean UV luminosity $\bar{L}_{\mathrm{UV}}$ within the absolute (AB)  magnitude system through the phenomenological relation~\cite{1983ApJ...266..713O}
    \begin{equation}
        0.4 \, \big(51.63 - \bar{M}_{\mathrm{UV}}\big) 
        = \log_{10}\!\left(\frac{\bar{L}_{\mathrm{UV}}}{\mathrm{erg\,s^{-1}}}\right)\,.
        \label{eq:luminosity}
    \end{equation}

    \item 
    Young stars emit strongly in the UV part of the electromagnetic spectrum, and stellar population synthesis models suggest a linear relation between the UV luminosity and the star formation rate (SFR)  $\dot{\bar{M}}_\star$:
    \begin{equation}
        \bar{L}_{\mathrm{UV}} = \frac{\dot{\bar{M}}_\star}{\kappa_\mathrm{UV}}\,,
        \label{eq:UV_lum}
    \end{equation}
    where $\kappa_\mathrm{UV} = 1.15 \times 10^{-28} \, M_\odot \,\mathrm{s\,erg^{-1}\,yr^{-1}}$ is a conversion factor calibrated from stellar population synthesis~\cite{Madau:2014bja}.
   \item 
    The SFR is then tied directly to the halo mass accretion rate:
    \begin{equation}
        \dot{\bar{M}}_{\star} = f_\star(M_h)\,\dot M_h\,.
        \label{eq:SFR_modelI}
    \end{equation}
    The efficiency function $f_*(M_h)$ is parameterized as a broken power law,
    \begin{equation}
         f_\star(M_h) = \frac{\epsilon_\star}{\left(\tfrac{M_h}{M_c}\right)^{\alpha_\star} + \left(\tfrac{M_h}{M_c}\right)^{\beta_\star}}\,,
        \label{eq:double_power_modelI}
    \end{equation}
 
    where $\epsilon_\star$ controls the amplitude, $M_c$ the characteristic SFR-peak mass, $\alpha_\star$ the low-mass slope, and $\beta_\star$ the high-mass slope of the UVLF. The parameters $\epsilon_\star$ and $M_c$ are allowed to be time-dependent, with a redshift evolution 
    modeled by the power-law fiducial choice of~\cite{Sabti:2021xvh}: 
    \begin{equation} \label{eq:epsilonz}
        \log_{10} \epsilon_\star(z) =   \epsilon_\star^s \times \log_{10}\left(\frac{1+z}{1+6}\right) + \epsilon_\star^i\,, 
    \end{equation}    
    \begin{equation} \label{eq:Masscz}
        \log_{10} M_c(z) =   M_c^s \times \log_{10}\left(\frac{1+z}{1+6}\right) + M_c^i\,. 
    \end{equation}
  
    \item 
    The halo mass accretion rate $\dot M_h$ entering Eq.\,\eqref{eq:SFR_modelI} is computed using the extended Press–Schechter formalism~\cite{Lacey:1993iv}, which leads to
    \begin{equation}
        \dot M_h(z) \;=\; \sqrt{\frac{2}{\pi}} \, 
        \frac{(1+z)H(z)M_h}{\sqrt{\sigma^2(M_h/Q) - \sigma^2(M_h)}} 
        \frac{\delta_{\rm cr}}{D^2(z)} \, \frac{dD(z)}{dz} \,,
        \label{eq:Mdot_h}
    \end{equation}
    where $D(z)=T_{\rm ad}(z)$ is the linear $\Lambda$CDM growth factor, $\sigma^2(M)$ the variance given in Eq.\,\eqref{eq:variance} (which includes both $\mathcal{P}_{\rm ad}$ and $\mathcal{P}_{\rm iso}$ of Eq.\,\eqref{eq:Piso1}), and $Q$ a free dimensionless parameter describing the step in halo mass used in the accretion calculation (which will be marginalized over). 

\begin{figure}[htbp]
    \centering
    \includegraphics[width=0.5\linewidth]{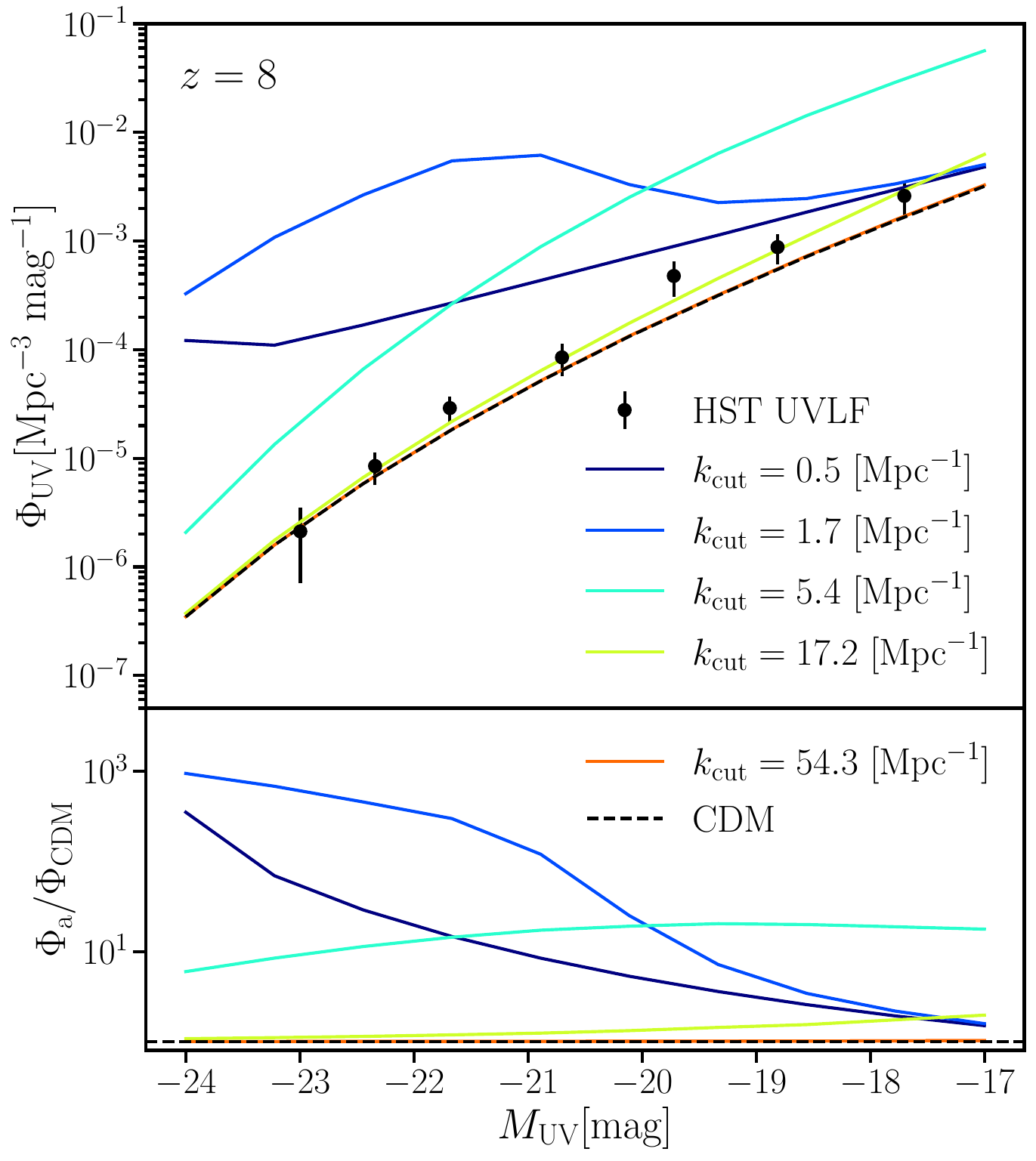}
    \caption{{\small UV luminosity function for standard CDM (black dashed) and for an additonal white-noise isocurvature component corresponding to varying values of its characteristic cutoff wavenumber, $k_{\rm cut}$, at redshift $z=8$. As in Fig.~\ref{fig:halmasscomparison}, the values $k_{\rm cut}=\{0.5,1.7,5.4,17.2,54.3\}$ $\rm Mpc^{-1}$ correspond to isocurvature components $\tilde{A}_{\rm iso}=\{1.2\times10^{-2},3.9\times10^{-4},1.2\times10^{-5},3.9\times10^{-7},1.2\times10^{-8}\}$ $\rm Mpc^{3}$, respectively. The black points correspond to the UVLF HST measurements at $z=8$, while the lower panels show the fractional difference with regards to the standard CDM prediction.}  }   \label{fig:UVLFcomparison} 
\end{figure}

We note that, strictly speaking, Eq.~\eqref{eq:Mdot_h} describes the halo accretion rate only within $\Lambda$CDM, since it employs the $\Lambda$CDM growth factor $D(z)$, even though the variance $\sigma^2(M_h)$ also includes the isocurvature contribution. 

However, $T_{\rm ad}$ and $T_{\rm iso}$ are not substantially different at the redshifts under consideration. In addition, Ref.~\cite{Sabti:2021xvh} demonstrated that the above method (called \emph{model I} in that work) reproduces an $M_h$-$M_{UV}$ relationship that is statistically identical with the ones produced by an empirical, observation-calibrated, model  (\emph{model III}) and a different, $\Lambda$CDM-independent, theoretical relationship as well (\emph{model II}).\footnote{The consistency between the results by models II $\&$ III was also confirmed in Ref.~\cite{Winch:2024mrt}. That analysis did not consider model I, as it focused on the regime in which the axion’s quantum pressure suppresses the transfer function, as described by $T_J$ in Eq. \eqref{eq:Pak} (see \cite{Poulin:2018dzj}), resulting in a growth factor much smaller than the $D(z)$ used in Eq.\,\eqref{eq:Mdot_h}.
} 
\end{enumerate}
In Fig.~\ref{fig:UVLFcomparison} we plot the UVLF function  for varying values of the cutoff wavenumber $k_{\rm cut}$ and $\tilde A_{\rm iso}$. As in Fig.~\ref{fig:halmasscomparison}, for fixed $A_{\rm iso}
$ (see Eq.~\ref{eq:Piso1}), lower $k_{\rm cut}$ values lead to increasing deviations from CDM, which are in tension with the HST UVLF data obtained in the same redshift bin ($z=8$). In Fig.~\ref{fig:PhiUV} we also plot the best-fit UVLF predictions obtained from our likelihood analysis (explained in the following section) alongside the HST data in the 7 redshifts bins we consider.

\begin{figure}[t]
    \centering
    \includegraphics[width=0.6\textwidth]{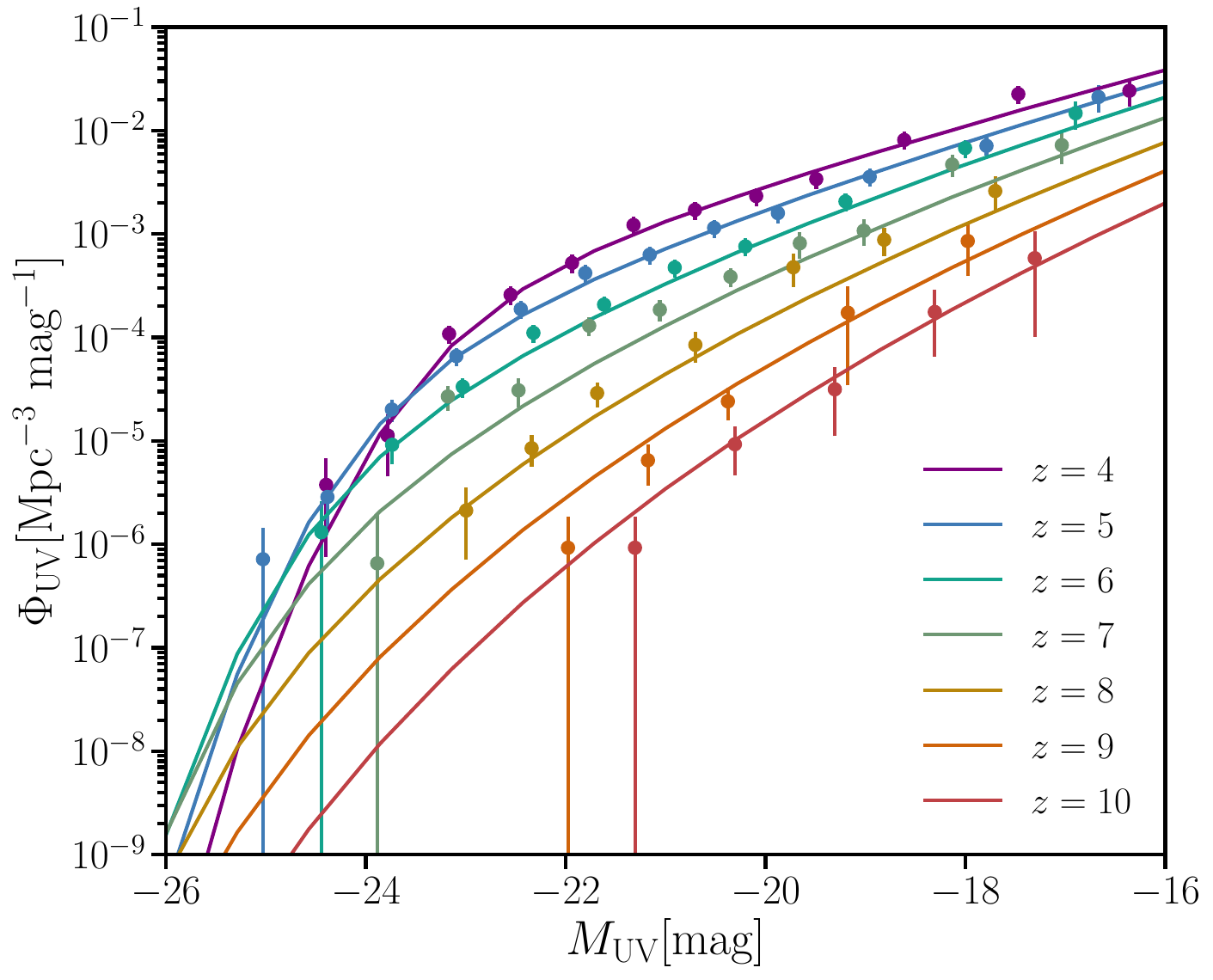}
    \caption{{\small The UVLF function data points from the HST are plotted in the redshift range $z=4-10$, together with the corresponding best-fit predictions from our likelihood analysis (solid lines). }}
    \label{fig:PhiUV}
\end{figure}

\vspace{3mm}
\noindent
\textbf{Likelihood analysis.}

\noindent
To constrain our models and, in particular, the isocurvature spectrum parameters $A_{\rm iso}$ and $k_{\rm cut}$, we perform a likelihood analysis of the UVLF data $(M_{\rm UV},\Phi_{\rm UV})$ obtained by the HST in seven redshift bins in the range $z=4-10$ (shown in Fig.~\ref{fig:PhiUV}) using the public package \texttt{GALLUMI}~\citep{Sabti:2021unj,Sabti:2021xvh}. The analysis in \texttt{GALLUMI} is performed using Markov Chain Monte Carlo runs based on the package \texttt{Montepython}~\citep{Audren:2012wb,Brinckmann:2018cvx}. For each point in the parameter space, the modified power spectrum is evaluated from Eq.~\eqref{eq:Pak}, with the adiabatic part given by the Boltzmann solver \textsc{CLASS}~\citep{2011arXiv1104.2932L}. Because each redshift bin in \texttt{GALLUMI} is analyzed separately, the isocurvature component is decomposed into the product of the time-dependent growth factor $T_{\rm iso}^2$ and the multiplicative offset $\tilde{A}_{\rm iso}$, which is the parameter actually constrained by the chains (for fixed $k_{\rm cut}$ values).

\begin{table}[t]
\centering
\setlength{\tabcolsep}{4pt}  
\renewcommand{\arraystretch}{1.1}  
\begin{tabular}{|c|c|c|c|}
\hline
\multicolumn{2}{|c|}{\textbf{Cosmological parameters}} &
\multicolumn{2}{c|}{\textbf{UVLF parameters}} \\
\hline
\textbf{Parameter} & \textbf{Prior} &
\textbf{Parameter} & \textbf{Prior} \\
\hline
$\omega_\mathrm{cdm}$  & $\mathcal{U}[-\infty,\,\infty]$ &
$\alpha_\star$ & $\mathcal{N}(-0.557,\,0.05)$ \\
$\omega_b$  & $\mathcal{U}[-\infty,\,\infty]$ &
$\beta_\star$ & $\mathcal{U}[0.0,\,3.0]$ \\
$A_s \times 10^9$ & $\mathcal{U}[-\infty,\,\infty]$ &
$\epsilon^\star_s$ & $\mathcal{U}[-3.0,\,3.0]$ \\
$n_s$ & $\mathcal{U}[-\infty,\,\infty]$ &
$\epsilon^\star_i$ & $\mathcal{U}[-3.0,\,3.0]$ \\
$h$ & $\mathcal{U}[-\infty,\,\infty]$ &
$M_c^s$ $[\rm M_{\odot}]$ & $\mathcal{U}[7.0,\,15.0]$ \\
$\tau$ & $\mathcal{U}[0.004,\,\infty]$ &
$M_c^i$ $[\rm M_{\odot}]$ & $\mathcal{U}[7.0,\,15.0]$ \\
$\tilde{A}_\mathrm{iso}\,[\rm Mpc^{3}]$ & $\mathcal{U}[0.0,\,10^{-4}]$ &
$\sigma_{M_{\rm UV}}$ $[\rm mag^{2}]$ & $\mathcal{U}[0.001,\,3.0]$ \\
 &  &
$Q$ & $\mathcal{U}[1.5,\,2.5]$ \\
\hline
\end{tabular}
\caption{{\small Priors choices adopted for the sampling parameters in our joint analysis using the HST UVLF and {\it Planck} CMB likelihoods.}}
\label{table:params}
\end{table}

Given a modified power spectrum, the UVLF function is predicted by Eqs.~\eqref{eq:radiusdef}-\eqref{eq:Mdot_h}. The complete set of parameters consists of the $\Lambda$CDM cosmological parameters, $(\omega_b, \omega_{cdm}, h, A_s, n_s)$, where $\omega_x = \Omega_{x} / h^2$ for the baryon and CDM density fraction and $h=H_{0}/100$ $\left(\rm km\cdot s^{-1} Mpc^{-1}\right)$ with $H_{0}$ the Hubble constant, the axion offset, $\tilde{A}_\mathrm{iso}$ and $k_{\rm cut}$ and the galaxy-halo connection parameters $(\epsilon^s_\star, \epsilon^i_\star, M^s_c, M^i_c,\alpha_\star, \beta_\star, \sigma_{UV}, Q)$ that are marginalized over. The parameter ranges and sampling strategy are listed in Table~\ref{table:params}. Convergence of our chains is determined by the Gelman-Rubin criterion, demanding $\tilde{R}-1<0.03$ for all parameters. In order to investigate how our constraints vary as a function of the cutoff wavenumber, we perform 8 separate runs for the following discrete values, $k_{\rm cut}=( 1, 2, 4, 6, 8, 10, 15, 20) \, \rm Mpc^{-1}$.  By combining our analysis with the CMB likelihood from {\it Planck} 2018 \cite{Planck:2018vyg}, we are able to exploit the full constraining power of the UVLF data spanning $7$ redshift bins in the range $4-10$. 

\begin{figure}
    \centering
    \includegraphics[width=0.9\linewidth]{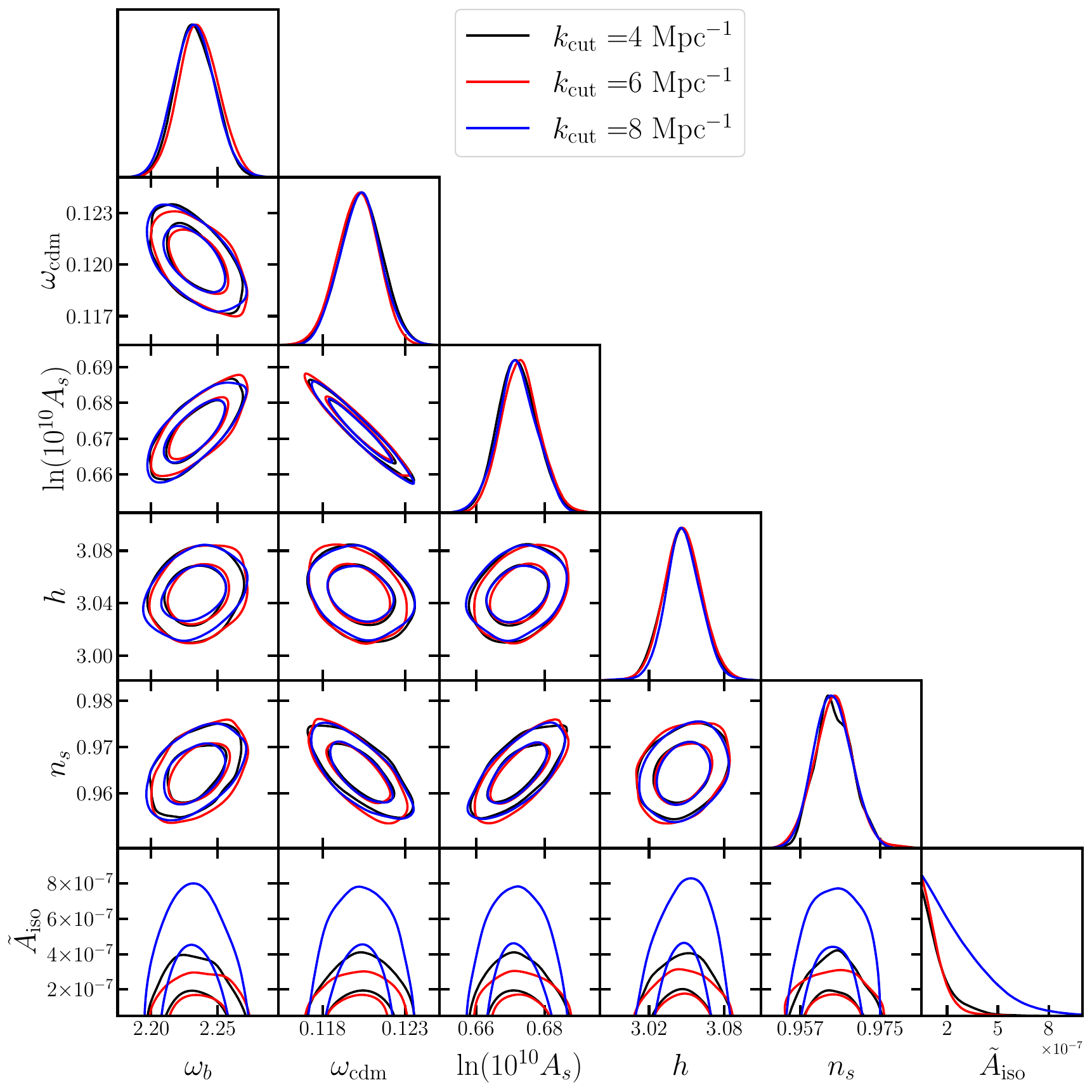}
    \caption{\small{Marginalized 68\% C.L. and 95\% C.L. contours obtained on the cosmological parameters and $\tilde{A}_\mathrm{iso}$ from our likelihood analysis performed for various cutoff wavenumbers, $k_{\rm cut}$. For visual clarity, only a subset of $k_{\rm cut}$ values is shown here; the full set of $\tilde{A}_\mathrm{iso}$ constraints is reported in Table~\ref{tab:Piso_bounds}.}}
    \label{fig:corner_cosmo}
\end{figure}

\begin{figure}
    \centering
    \includegraphics[width=0.98\linewidth]{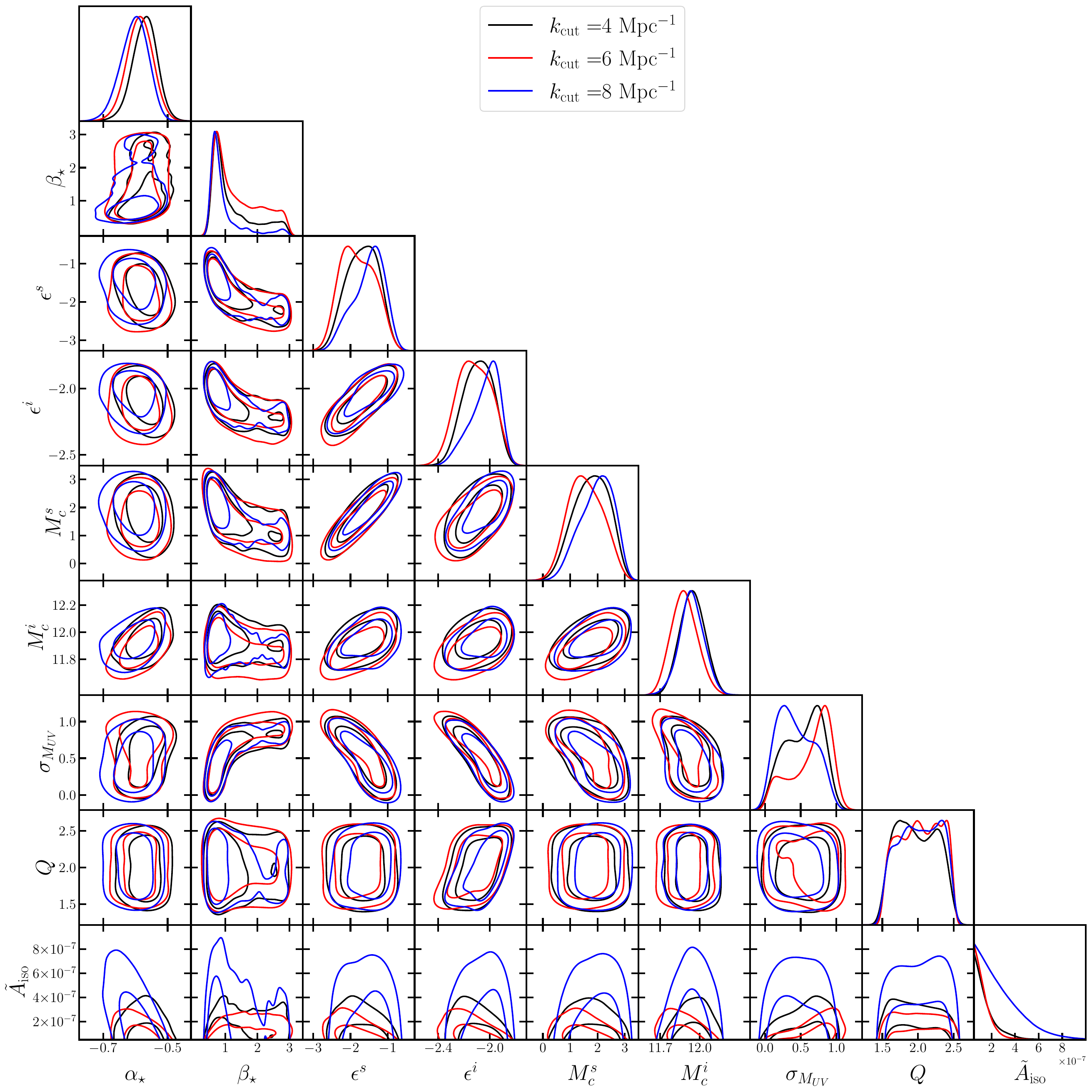}
    \caption{\small{Marginalized 68\% C.L. and 95\% C.L. contours obtained on the UVLF galaxy parameters and $\tilde{A}_\mathrm{iso}$ from our likelihood analysis performed for various cutoff wavenumbers, $k_{\rm cut}$.}}
    \label{fig:corner_hod}
\end{figure}

In Fig.~\ref{fig:corner_cosmo}, we plot the marginalized 68\% C.L. and 95\% C.L. contours obtained on the cosmological parameters from the joint UVLF+{\it Planck} likelihood analysis for each one of the cutoff wavenumbers $k_{\rm cut}$. The corresponding $68\%$ $\&$ $95\%$ upper limits on the axion-induced $\tilde{A}_\mathrm{iso}$ for each case are listed in Table~\ref{tab:Piso_bounds} and illustrated in Fig.~\ref{fig:Piso} (green region). Our analysis finds no preference for $\tilde{A}_\mathrm{iso}$, which is constrained to be consistent with zero for all $k_{\rm cut}$ cases. As can be seen from Table~\ref{tab:Piso_bounds} and Fig.~\ref{fig:Ptot}, the upper bounds we obtain on $\tilde{A}_\mathrm{iso}$ overall get increasingly tighter as we allow larger $k_{\rm cut}$ values, as expected. The trend is not perfectly monotonic, however, due to internal degeneracies between $\tilde{A}_\mathrm{iso}$ and the UVLF model galaxy parameters (see discussion below), the constraints of which are shown in Fig.~\ref{fig:corner_hod}. 

It is worth mentioning that our analysis found a strong degeneracy between $\tilde{A}_\mathrm{iso}$ and the faint-end slope parameter $\alpha_{\star}$, which can be explained as follows: lower (more negative) values of $\alpha_{\star}$ lead to a suppression of the UVLF (see Fig. 2 in Ref.~\citep{Sabti:2021xvh}), which can counteract the power enhancement caused by a positive $\tilde{A}_\mathrm{iso}$ and weaken the constraining power of the UVLF. To overcome this issue, we impose (as we show in Table~\ref{table:params}) a more informative Gaussian prior on $\alpha_{\star}$ based on the constraints obtained on it from the $\Lambda$CDM analysis of Ref.~\citep{Sabti:2021unj}, which allows us to robustly constrain deviations in the form of $\tilde{A}_\mathrm{iso}$. Furthermore, the bright-end slope parameter $\beta_{\star}$ is found to be unconstrained at the 95\% C.L. level for all $k_{\mathrm{cut}}$ cases, which is consistent with the findings of the $\Lambda$CDM UVLF analysis of Ref.~\citep{Sabti:2021xvh}. As further discussed in~\citep{Sabti:2021xvh}, 
$\beta_{\star}$ primarily governs the bright-end tail of the UVLF, where Poisson uncertainties dominate. Consequently, the data do not provide meaningful constraints on this parameter. The features that appear in the contour plot for $\beta_\star$ are related to the fact that, since this parameter is unconstrained, the results are more noisy at the edge of the prior space. Our constraints are overall further tightened by the joint analysis with the {\it Planck} likelihood, which strongly constrains the $\Lambda$CDM parameters (including amplitude parameters, such as $A_s$) and reduces further degeneracies, as seen in Fig.~\ref{fig:corner_cosmo}.  
In Sec. \ref{sec:results} we will demonstrate how our results for $\tilde{A}_\mathrm{iso}$ can be translated to constraints on axion mass and decay constant in the post-inflationary scenario.
 
\begin{table}[t]
\centering
\begin{tabular}{|c|c|c|}
\hline
 & \multicolumn{2}{c|}{Constraints on $\tilde{A}_{\rm iso}\,[\rm Mpc^{-3}]$} \\
\hline
$k_{\rm cut}$ [Mpc$^{-1}$] & 68\% bound & 95\% bound \\
\hline
1   & $< \, 8.06\cdot 10^{-6}$ & $< \, 1.81\cdot 10^{-5}$ \\
2   & $< \, 8.8\cdot 10^{-7}$ & $< \, 1.93\cdot 10^{-6}$ \\
4   & $< \, 1.26\cdot 10^{-7}$ & $< \, 3.21\cdot 10^{-7}$ \\
6   & $< \, 1.12\cdot 10^{-7}$ & $< \, 2.42\cdot 10^{-7}$ \\
8   & $< \, 2.92\cdot 10^{-7}$ & $< \, 6.35\cdot 10^{-7}$ \\
10  & $< \, 3.08\cdot 10^{-7}$ & $< \, 6.33\cdot 10^{-7}$ \\
15  & $< \, 2.95\cdot 10^{-7}$ & $< \, 6.11\cdot 10^{-7}$ \\
20  & $< \, 2.91\cdot 10^{-7}$ & $< \, 6.20\cdot 10^{-7}$ \\
\hline
\end{tabular}
\caption{Marginalized constraints obtained on the primordial amplitude $\tilde{A}_{\rm iso}$ of the isocurvature power spectrum at 68\% and 95\% confidence levels for the various cutoff wavenumbers $k_{\rm cut}$.}
\label{tab:Piso_bounds}
\end{table}

\subsection{Other probes}
\label{sec:other}

{\bf Lyman-$\alpha$ forest.}\nopagebreak

\noindent
The Lyman-$\alpha$ forest provides another powerful probe of the matter distribution on comoving scales of a few Mpc at redshifts up to $z \sim 5$. It arises from a series of absorption lines in the spectra of distant quasars, produced by intervening neutral hydrogen in the intergalactic medium. Since the gas closely follows the underlying dark matter distribution, the Lyman-$\alpha$ flux power spectrum is highly sensitive to modifications of the matter power spectrum. Two complementary datasets are available: the eBOSS data~\cite{eBOSS:2018qyj,Chabanier:2019eai}, which covers redshifts $z \simeq 2.1$--$4.5$, and the high-resolution MIKE/HIRES quasar spectra, which extend up to $z \simeq 5.4$~\cite{Irsic:2017yje}.

Refs.~\cite{Murgia:2019duy,Ivanov:2025pbu} have used the high-redshift data to constrain the possibility that dark matter consists of supermassive primordial black holes. In that case, the discreteness of the primordial black holes distribution gives rise to an additional white-noise isocurvature contribution to the matter power spectrum. The resulting analyses translate into limits on the allowed primordial black hole abundance and mass range. Crucially for our purposes, the underlying statistical form of the white-noise is identical to that produced for post-inflationary axions (see Eq.\,\eqref{eq:Piso}). This allows us to directly reinterpret the existing bounds as constraints on $A_{\rm iso}$ as illustrated in Fig.~\ref{fig:Piso}. 

Specifically, Ref.~\cite{Ivanov:2025pbu} analyzed the MIKE/HIRES data within an effective field theory (EFT) framework for the one-dimensional flux power spectrum. This perturbative description captures gravitational non-linearities and redshift-space distortions, while marginalizing over EFT counterterms that parametrize the impact of unresolved small-scale astrophysics. A likelihood analysis of the data then yields the most stringent available constraints (blue region in Fig.~\ref{fig:Piso}) on white-noise type isocurvature fluctuations (see Eq.\,(7) of that paper). The EFT UV cutoff scale is set by the Lyman-$\alpha$ forest gas scale, while the lowest IR value accounts for the loss of data sensitivity at large scales. 

Additionally, Ref.~\cite{Murgia:2019duy} analyzed the same MIKE/HIRES spectra relying instead on a suite of hydrodynamic simulations including an isocurvature component in the initial conditions. This method can access the fully non-linear regime where effective field theory breaks down, but at the same time it is limited to relatively large isocurvature amplitudes, since the white-noise description of the primordial black hole population becomes unreliable for low abundances~\cite{Carr:2018rid}. As a consequence, the simulation-driven bounds on $\mathcal{P}_{\rm iso}$ are somewhat more restrictive in coverage in terms of large-scales.  To translate the bounds on $A_{\rm iso}$, Ref.~\cite{Murgia:2019duy} provides the limit on the primordial isocurvature-to-adiabatic amplitude ratio $f_{\rm iso}\equiv \sqrt{\mathcal{P}_{\rm iso}/\mathcal{P}_{\rm ad}}=\sqrt{(A_{\rm iso}/A_s)(k_p/k_{\rm cut})^3}\lesssim 0.004$ at the reference CMB pivot scale $k_p=0.05\,{\rm Mpc}^{-1}$ (derived using $M_{\rm PBH} \lesssim 170\,M_\odot$ for $f_{\rm DM}=1$), which leads to the red region in Fig.~\ref{fig:Piso}. The IR cutoff ($k\sim 50\,{\rm Mpc}^{-1}$) is obtained by calculating the cutoff scale of $\mathcal{P}_{\rm iso}$ in primordial black hole models (see Eq.\,(5) in Ref.~\cite{Inman:2019wvr}) using the maximum mass and fraction products that were included in the simulations of Ref.~\cite{Murgia:2019duy}, and it reflects the fact that it is not clear how to extrapolate the bound to higher masses in this framework~\cite{Carr:2020gox}.

\emph{Fuzzy dark matter.} 
As discussed in Sec.\,\ref{sec:theory}, quantum pressure suppresses the formation of structures for $k>k_j^{\rm eq}$ irrespective of the additional contribution $\mathcal{P}_{\rm iso}$. Refs.~\cite{Kobayashi:2017jcf,Irsic:2017yje} derived constraints on $m_a$ and $f_{\rm DM}$ from this effect by modeling the Lyman-$\alpha$ flux power spectrum using full hydrodynamical simulations. Comparing the simulated flux spectra with the XQ-100 and HIRES/MIKE quasar samples, Refs.~\cite{Irsic:2017yje} obtained a lower bound on the particle mass of $m_a > 2\times10^{-21}\,{\rm eV}$ (95\% C.L.) for the combined dataset, tightening to $m_a > 3.8\times10^{-21}\,{\rm eV}$ under smoother thermal-history assumptions. These results correspond to an upper limit on the suppression scale of the matter power spectrum at $k\lesssim10\,{\rm Mpc}^{-1}$. In addition, Ref.~\cite{Kobayashi:2017jcf} obtained a more general bound for $f_{\rm DM}<1$, reported in Fig.\,\ref{fig:money_plot} for the axion.

\vspace{3mm}
\noindent
{\bf CMB/BAO bounds.}\nopagebreak

\noindent
Ref.~\cite{Buckley:2025zgh} derived constraints on isocurvature fluctuations by analyzing CMB temperature, polarization, and lensing measurements from Planck 2018~\cite{Planck:2018vyg} together with BAO distance measurements from BOSS DR12~\cite{BOSS:2016wmc}. Their study employed a broken power-law parameterization of $\mathcal{P}_{\rm iso}$ that mimics the white-noise form in Eqs.\,\eqref{eq:Piso} and \eqref{eq:Piso1}. An almost scale-invariant contribution to the gravitational potential across the CMB window  alters the relative heights and phases of the acoustic peaks in the temperature and polarization spectra and shifts the inferred BAO distance scale. 
They find that the constraints are driven by sensitivity near the CMB pivot scale. For scales below this value, the limits on the isocurvature amplitude are essentially flat, and as noted in Ref.~\cite{Gerlach:2025vco}, the bottom left plot in Fig. 4 in Ref.~\cite{Buckley:2025zgh} describes well the bounds on the form of $\mathcal P_{\rm iso}$ we use in this work. We report these bounds in Fig. \ref{fig:Piso} (yellow region). In practice, $A_{\rm iso}$ above a few times $10^{-10}$ are excluded at 95\% confidence. Similar magnitude constraints on white-noise isocurvature spectra from CMB observations (valid for $k\gtrsim k_p$, but not shown in Fig. \ref{fig:Piso}) have also been obtained in Refs.\,\cite{Feix:2020txt,Feix:2019lpo}.

\section{Implication on axion mass and decay constant}
\label{sec:results}

\begin{figure}[t]
    \ \ \includegraphics[width=.93\textwidth]{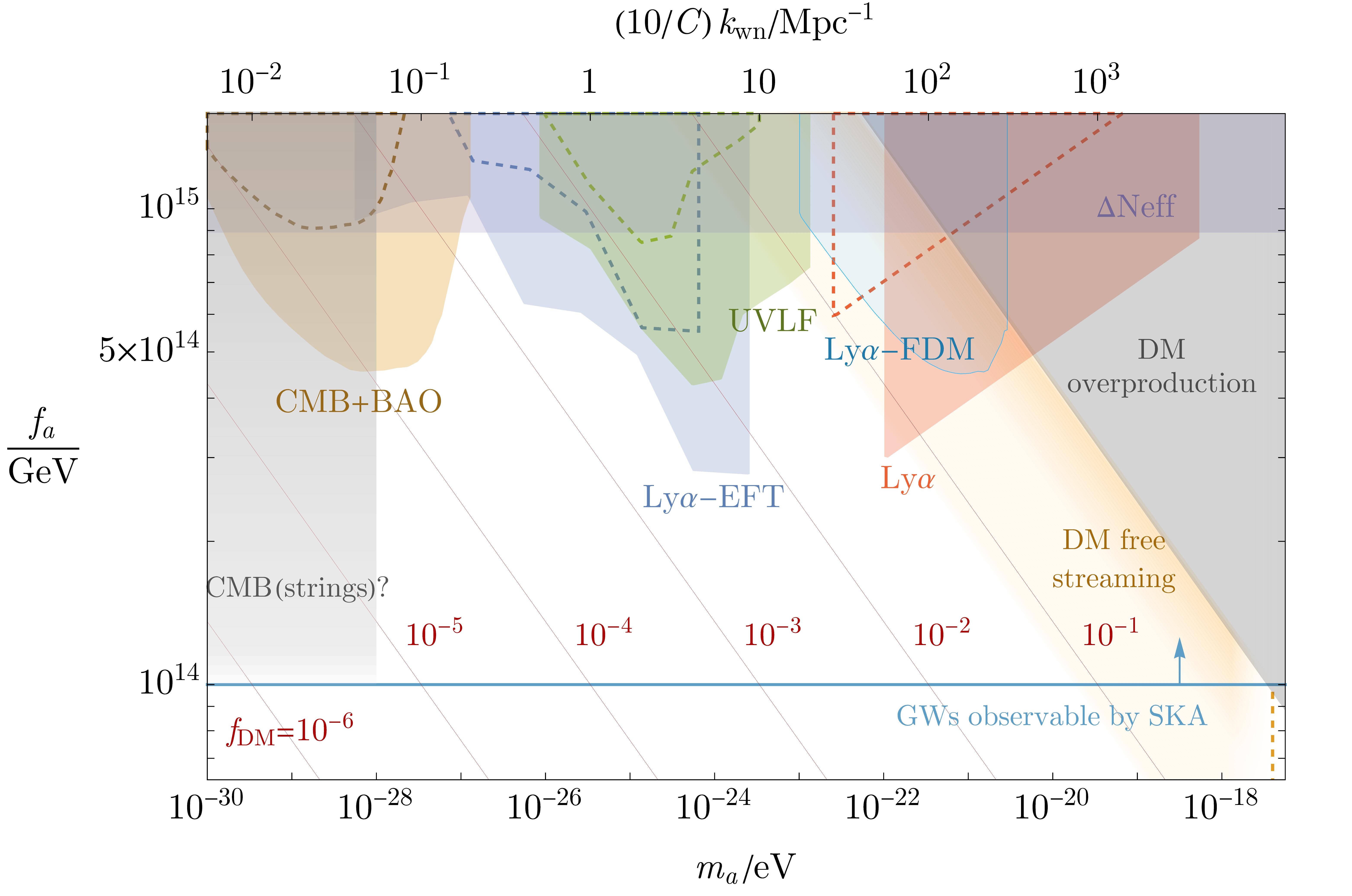}
    \caption{ 
    {\small 
    Constraints on the axion mass $m_a$ and decay constant $f_a$ in the post-inflationary scenario, including large-scale structure probes. Axions radiated by strings can overproduce dark matter (gray) or contribute as dark radiation $\Delta N_{\rm eff}$ (purple). CMB observations exclude very light $m_a$ at large $f_a$ via string-induced anisotropies (``CMB strings'', gray). Thin red lines indicate $(m_a,f_a)$ for which axions make up a fraction $f_{\rm DM}$ of dark matter. For $f_{\rm DM}\simeq1$, free-streaming excludes sufficiently small $m_a$ (yellow). For smaller $f_{\rm DM}$, the most stringent bounds arise from the enhancement of CDM growth sourced by the axion white-noise power (CMB+BAO, Lyman-$\alpha$, and galaxy UVLF; see Fig.~\ref{fig:Piso}) and are presented for the benchmark $C=10$ (shaded regions). The same bounds are also shown for $C=20$ (dashed), illustrating the size of the theoretical uncertainty.} The top axis shows the white-noise peak scale $k_{\rm wn}=C\,k_\star$.}
    \label{fig:money_plot}
\end{figure}

In this section we combine the LSS bounds discussed in Sec.~\ref{sec:LSS} with other probes of post-inflationary axions. The final results are summarized in Fig.~\ref{fig:money_plot}, which presents the combined and most up-to-date constraints on the axion mass $m_a$ and decay constant $f_a$.  

\vspace{1mm}
\noindent
{\bf Dark matter.}  Axions with masses $m_a \lesssim \mathrm{MeV}$ are stable on cosmological times (see e.g. Refs.~\cite{Arias:2012az,Gorghetto:2021fsn}), and  
can overproduce the observed dark matter abundance if $f_a$ or $m_a$ are large enough. We show in gray in Fig.~\ref{fig:money_plot} the bound $\Omega_a< \Omega_{\mathrm{DM}}$ from dark matter overproduction, where $\Omega_a$ is approximated by the relic axions from the string scaling regime as in Eqs.~\eqref{eq:relic} and~\eqref{eq:Omegast}; we assume $x_0 = 10$, $m_r\simeq f_a$ and that $\xi\simeq 0.24\log(m_r/H)$ also at $\log(m_r/H) \gg1$. We also assume that the string network persists until $H = m_a/3$, as suggested by the simulations for a temperature-independent axion mass in Ref.~\cite{Gorghetto:2021fsn}, which increases the abundance in Eq.~\eqref{eq:relic} by a factor $\sqrt{3}$. However, as emphasized in Sec.~\ref{sec:theory}, significant uncertainties remain in the determination of the abundance. The thin red lines Fig.~\ref{fig:money_plot} indicate the values of $m_a$ and $f_a$ that corresponds to fixed values of $f_{\rm DM}$. 

\vspace{2mm}
\noindent
{\bf Axion dark radiation.}   
The axions emitted during the string scaling regime that remain relativistic until Big-Bang Nucleosynthesis (BBN) or CMB decoupling contribute to dark radiation, with energy density $\rho_{\rm rel}$, and are potentially observable~\cite{Gorghetto:2021fsn}. 
Constraints on $\rho_{\rm rel}$ are  expressed in terms of the effective number of neutrino species, 
\begin{equation}
    \Delta N_{\rm eff} = \frac{8}{7}\left(\frac{11}{4}\right)^{4/3} \frac{\rho_{\rm rel}}{\rho_\gamma} \, ,
\end{equation}
where $\rho_\gamma$ is the photon energy density. 
The relativistic axion energy density can be estimated as
\begin{equation}
    \rho_{\rm rel}(t) 
      = \int_{t_0}^{t} dt'\, \Gamma^{\rm st}(t') \left[\frac{a(t')}{a(t)}\right]^4
      \;\simeq\; \frac{4}{3}\pi c_1 f_a^2 \log^3\!\left(\frac{f_a}{H(t)}\right)\,,
\end{equation}
where $c_1 \simeq 0.24$, and the dependence on the initial time $t_0$ is weak. 
Thus $\rho_{\rm rel}$ depends primarily on the decay constant $f_a$, which controls the string tension and hence the axion emission rate $\Gamma^{\rm st}$.  
Current $95\%$ C.L. limits on $\Delta N_{\rm eff}$ from BBN and the CMB by Planck~\cite{Planck:2018vyg,Pitrou:2018cgg}, $\Delta N_{\rm eff}\lesssim 0.28$, imply $f_a \lesssim 10^{15}\,\GeV$, shown in purple in Fig.~\ref{fig:money_plot}. 
Future improvements in $\Delta N_{\rm eff}$ sensitivity to $\lesssim 0.04$~\cite{Abazajian:2013oma} could strengthen this bound to $f_a \lesssim 2\times 10^{14}\,\GeV$.

\vspace{2mm}
\noindent
{\bf LSS bounds.}   

\noindent {\bf \emph{Free streaming.}} If the axion makes all of the dark matter, the most IR modification of the matter power spectrum arises from the free-streaming suppression of adiabatic perturbations discussed in Sec.~\ref{sec:scales}, which imposes the strongest limit on $m_a$. Requiring that the free-streaming momentum at matter-radiation equality, $k_{\rm fs} \simeq k_\star/[C\,\log(a_{\rm eq}/a_\star)]$, see Eq.\,\eqref{eq:kfs}, is larger than the largest observable momentum, roughly $10~{\rm Mpc}^{-1}$, yields the bound $m_a\gtrsim 10^{-18}$\,eV for $f_{\rm DM}=1$, where we used conservatively $C=10$. This excludes the yellow region in Fig.~\ref{fig:money_plot}.

This bound inherits the large uncertainties associated with the power spectrum, ultimately stemming from $C$—more precisely, from the typical momentum $k_a$ in the spectrum of Fig.~\ref{fig:Piso-sim} (left)—as well as those related to the dark matter abundance. Since $k_\star\propto m_a^{1/2}$, a higher peak would translate into a quadratically stronger bound on $m_a$. For example, $C= 20$ yields $m_a\gtrsim 4\times10^{-18}\,\mathrm{eV}$, shown for reference by the dashed yellow line in Fig.~\ref{fig:money_plot}.

For $f_{\rm DM} < 1$, this constraint gradually weakens. As shown in Eq.\,\eqref{eq:T2ad}, free-streaming reduces the adiabatic power by a factor $(1-f_{\rm DM})^2$ at the scale $k\simeq k_{\rm fs}=k_\star/(C\log(a_{\rm eq}/a_\star))\simeq 0.5\,{\rm Mpc}^{-1}(m_a/10^{-20}\eV)^{1/2}$. Imposing that the adiabatic piece does not deviate from the $\Lambda$CDM expectation by more than $10\%$ we obtain a bound on $f_{\rm DM}\lesssim0.05$, as illustrated by the shaded yellow region. A precise analysis of this effect is left for future work.
\vspace{1mm}

\noindent {\bf \emph{Isocurvature fluctuations.}} As discussed in Sec.~\ref{sec:scales} and Fig.~\ref{fig:k12}, if $f_{\rm DM}=1$ the axion evolves as CDM for $k\lesssim k_J^{\rm eq}$, while its spectrum is suppressed for  $k\gtrsim k_J^{\rm eq}$. For $f_{\rm DM}\ll1$, effects of velocity dispersion suppress the CDM power spectrum by a factor $k_J^{\rm eq}/k$, see Eq.\,\eqref{eq:TisoJ}. Because the mixed axion+CDM system does not evolve like pure CDM, the white-noise bound on enhanced structure formation should be applied with care. As anticipated in Sec.~\ref{sec:LSS}, one can, first, derive LSS bounds by adopting the choice $k_{\rm cut}=k_J^{\rm eq}$. For the benchmark $C=10$, the resulting limits are generally weaker than those from dark radiation ($\Delta N_{\rm eff}$), except for the Lyman-$\alpha$ EFT constraint (not shown in Fig.\,\ref{fig:money_plot} to avoid clutter).

As discussed in Sec.~\ref{sec:LSS}, for $f_{\rm DM}\ll 1$, stronger bounds can be obtained by accounting for the structure formation induced by the CDM component from higher-amplitude modes with $k>k_J^{\rm eq}$ taking $k_{\rm cut}=k_{\rm wn}$ and rescaling the isocurvature amplitude by $k_J^{\rm eq}/k_{\rm wn}\simeq 2C^{-2}$. This tighter treatment is less rigorous for early-time probes such as CMB+BAO, but it remains well motivated for late-time observables like the UVLF and the Lyman-$\alpha$ forest. The bounds resulting from this are presented below.

The orange region (CMB+BAO) shows the limit from the axion-induced dark matter isocurvature fluctuations constrained by the CMB anisotropies and the BAO distance scale. These measurements are most sensitive to modes $k_{\rm wn}$ of order the Hubble parameter at decoupling (see Fig.~\ref{fig:Piso}), corresponding roughly to
$m_a\simeq 10^{-28}-10^{-29}$\,eV for $C=10$. The bounds exclude high decay constants, $f_a \gtrsim 5\cdot10^{14}\,{\rm GeV}$. Ultimately, the exclusion arises because larger $f_a$ enhances the string tension and thus the energy emitted into axions, leading to an excessive isocurvature contribution to $f_{\rm DM}$ that would distort the precisely measured CMB acoustic pattern.  

At smaller scales, the Lyman-$\alpha$ bound based on the EFT analysis (Ly$\alpha$-EFT, blue) and the UVLF constraint (UVLF, green) test the bulk of the parameter space.
Among these, the Lyman-$\alpha$ EFT probes  momenta $k_{\rm wn}\simeq 0.05-10\, {\rm Mpc}^{-1}$, corresponding approximately to  $m_a\simeq 10^{-28}-10^{-24}\,{\rm eV}$, and are the most stringent, excluding values of $f_a$ down to $5\times10^{14}\,{\rm GeV}$. The UVLF analysis, on the other hand, probes $k_{\rm wn}\simeq 1-20\, {\rm Mpc}^{-1}$ where the abundance of early (smaller) galaxies is most sensitive to enhanced small-scale power. These scales corresponds to $m_a\simeq 10^{-27}-10^{-23}\,{\rm eV}$, delivering the leading bounds in the narrower window $m_a\simeq 10^{-24}-10^{-23}\,{\rm eV}$. 
Both constraints display cutoffs at low and high $m_a$ (which depend on $C$), corresponding to the IR and UV limits in Fig.~\ref{fig:Piso}, where the white-noise momentum peak falls below or above the scales probed by the analyses.
The Lyman-$\alpha$ bound based on hydrodynamic simulations (Ly$\alpha$, red) instead tests somewhat higher momenta ($k_{\rm wn}\gtrsim 50\,{\rm Mpc}^{-1}$) and excludes $m_a\gtrsim10^{-22}\,\eV$.   

Similarly to the free-streaming bound, the limits from isocurvature fluctuations are affected by theoretical uncertainties, arising from the poorly known amplitude of the white-noise tail and the average axion momentum, collectively parametrized by $C$ (see Sec.~\ref{sec:theory}). To illustrate their impact, Fig.~\ref{fig:money_plot} also shows with dashed lines the bounds for $C=20$. For such more UV spectra the isocurvature bounds are weaker, as the number of halos is reduced due to both stronger free streaming and a smaller isocurvature tail amplitude. This behavior is reflected in the scaling $k_J^{\rm eq}/k_{\rm wn}\simeq 2C^{-2}$. In any case, as long as $C\lesssim 30$, the Lyman-$\alpha$ EFT bound remains stronger than the limits from $\Delta N_{\rm eff}$.

We stress that, if domain walls dominate the abundance, $C$ could be smaller (even $C\simeq \mathcal O(1)$), leading to substantially stronger bounds from enhanced structures. In this case the abundance may also change considerably (see Sec.~\ref{sec:theory}), which would shift the `DM overproduction' region in Fig.~\ref{fig:money_plot}. Since neither the abundance nor the spectrum in this scenario has been computed in simulations with realistic parameters, nor reliably extrapolated, we do not show these bounds.

\vspace{1mm}
\noindent
{\bf \emph{Quantum Jeans suppression.}} Finally, a light blue contour (Ly$\alpha-$FDM) indicates the bound on the fraction of dark matter that can be fuzzy, taken from Ref.~\cite{Kobayashi:2017jcf} and discussed in Sec.~\ref{sec:other}. More recent analyses from the HST UVLF, Planck CMB and BOSS galaxy surveys provide slightly more restrictive limits~\cite{Rogers:2023ezo,Winch:2024mrt}. 
Unlike the bounds discussed above, all of these do not depend on our knowledge of $\mathcal{P}_{\rm iso}$ and are thus more robust.

\vspace{2mm}
\noindent
{\bf Strings and CMB anisotropies.}   If strings persist at decoupling, their gravitational interactions source additional CMB anisotropies: i) The conical spacetime around a string induces a deficit angle, producing discontinuous temperature fluctuations in the CMB photons of order $\delta T / T \propto G\mu$~\cite{Zeldovich:1980gh,Vilenkin:1981iu,Kaiser:1984iv}. ii) A moving string in the primordial plasma generates a wake via purely gravitational interactions, leading to additional, incoherent density perturbations that do not manifest as BAO~\cite{Albrecht:1995bg,Magueijo:1995xj,Pen:1997ae,Pogosian:2003mz,Wyman:2005tu,Fraisse:2006xc,Ade:2013xla}. A comparison between these signatures and CMB observations yields an upper bound on the string tension $\mu$, and thus on $f_a$, for a string network that survives after decoupling, which occurs for ultra-light axions with $m_a \lesssim 10^{-28}\,\mathrm{eV}$ (Hubble at decoupling). Based on previous studies (e.g. Refs. \cite{Battye:2010xz,Charnock:2016nzm,Lizarraga:2016onn} for local strings and Ref. \cite{Lopez-Eiguren:2017dmc} for global strings), a bound of $G\mu \lesssim 10^{-7}$ is plausible, corresponding to $f_a \lesssim 2\times10^{14} \ \mathrm{GeV}$, shown in gray and labeled as \emph{CMB (strings)?} in Fig.~\ref{fig:money_plot}. Nevertheless, significant uncertainties remain, stemming from the lack of a conclusive extrapolation of the string network dynamics, which we indicate by the blurred boundary.

\vspace{2mm}
\noindent
{\bf Gravitational waves.}  
Finally, the axion string network acts as a continuous source of gravitational waves (GWs) once it enters the scaling regime. The emission arises from the oscillations and reconnections of long strings and loops, producing a nearly scale-invariant stochastic GW background over a wide frequency range, from nanohertz to millihertz. As shown in Ref.~\cite{Gorghetto:2021fsn,Gouttenoire:2019kij}, the amplitude of this background scales steeply with the string tension $\mu=\pi f_a^2\log(m_r/H)$ and the number of strings per Hubble patch $\xi$ leading to the GW energy density at frequency $f$: $\Omega_{\rm GW}(f) \propto f_a^4\log^4(f_a/f)$, which therefore grows rapidly with the axion decay constant. This makes gravitational waves an exceptionally clean and sensitive probe of high-$f_a$ models.  

Current pulsar-timing observations from the 15-year NANOGrav dataset~\cite{NANOGrav:2023gor} already constrain $f_a \lesssim 3\times10^{15}\,{\rm GeV}$, a bound weaker than that derived from dark-radiation measurements. However, future radio interferometers such as Square Kilometre Array (SKA)~\cite{Janssen:2014dka} are expected to improve this sensitivity by about an order of magnitude, reaching $f_a \simeq  10^{14}\,{\rm GeV}$ and potentially closing the remaining portion of parameter space below this scale, see blue line in Fig.~\ref{fig:money_plot}.

\vspace{2mm}
\noindent
{\bf Domain wall number \boldsymbol{$N > 1$}.}
Finally, we briefly discuss cosmological bounds in the $N > 1$ case, where, besides $m_a$ and $f_a$, the Hubble scale at domain-wall decay, $H_d$, is an additional free parameter. A detailed study is not attempted here due to sizable theoretical uncertainties and the lack of simulations at the relevant separation between $H_d$ and $m_a$. Many of these bounds have been highlighted in Refs.~\cite{Gorghetto:2022sue,Petrossian-Byrne:2025mto,Gelmini:2021yzu,Gelmini:2022nim}. 
As discussed in Sec.\,\ref{sec:theory}, axions are expected to be produced at $a = a_d$ with comoving momentum $k_a \simeq a_d m_a$. In contrast to the $N = 1$ case, the white-noise spectrum $\mathcal{P}_{\mathrm{iso}}$ peaks at a more infrared scale, $k_{\mathrm{dw}} = C_d a_d H_d \simeq C_d \cdot 54 (H_d / 10^{-20}\,{\rm eV})^{1/2} \, {\rm Mpc}^{-1}$.

The free streaming, quantum and classical Jeans momenta are  given by Eqs.\,\eqref{eq:kfs} and~\eqref{eq:kJs} with $k_J=k_j^2/(2k_a)$. Relative to the white-noise peak, they read
\begin{equation}\label{eq:kN>1}
    \frac{k_{\rm fs}^{\rm eq}}{k_{\rm dw}}\simeq \frac{1}{C_d\log(a_{\rm eq}/a_d)} \ , \qquad\quad \frac{k_{j }^{\rm eq}}{k_{\rm dw}}\simeq \frac{1}{C_d}\left(\frac{m_a}{H_d}\right)^\frac12 \, , \quad\qquad  \frac{k_{J}^{\rm eq}}{k_{\rm dw}}\simeq\frac{1}{C_d} \, ,
\end{equation}

Thus, in the case $H_d\ll m_a$ that we consider, the power spectrum is similar to Fig.~\ref{fig:Psketch}, but
quantum pressure is irrelevant in the evolution of the white-noise fluctuations and the velocity dispersion is marginally relevant (depending on whether $C_d\simeq 1$ or larger). On the other hand, the free-streaming momentum is a still factor $\sim (10C_d)^{-1}$ smaller than $k_{\rm dw}$.  
Free-streaming dumps adiabatic perturbations (as well as the peak of $\mathcal{P}_{\rm iso}$). Requiring as before the approximate observational limit $k_{\rm fs}^{\rm eq}\gtrsim10\,{\rm Mpc}^{-1}$ leads to $H_d\gtrsim 6\cdot 10^{-20}\,$eV for $f_{\rm DM}=1$, corresponding to $T_d\gtrsim10$\,keV.\footnote{This bound has been already noted in Refs.~\cite{Gelmini:2021yzu,Gelmini:2022nim} by analogy with the minimum warm dark matter mass.} In other words, if the domain network decays too late, the warm axions produced do not have sufficient time to cool, and free-stream over observable distances.

In addition, the white-noise fluctuations behave as CDM below the effective cutoff $k_{\rm cut}\simeq k_{\rm fs}^{\rm eq}$ for $f_{\rm DM}=1$, which, for the values of $H_d$ constrained by free-streaming above, is within the regime of the Lyman-$\alpha$ bound in Fig.~\ref{fig:Piso}. This bound is  $f_{\rm iso} =\sqrt{\mathcal{P}_{\rm iso}/\mathcal{P}_{\rm ad}}=\sqrt{(A_{\rm iso}/A_s)(k_p/k_{\rm cut})^3}=f_{\rm DM}\sqrt{(k_p/k_{\rm dw})^3/A_s} \lesssim0.004$  at the pivot scale $k_p=0.05\,{\rm Mpc}^{-1}$, which yields $H_d\gtrsim 7 \cdot 10^{-18}\eV f_{\rm DM}^{4/3}/C_d^2$, corresponding to $T_d\gtrsim100 \keV f_{\rm DM}^{2/3}/C_d$. 
If $C_d$ is not larger than $\mathcal{O}(10)$, this bound is stronger than that from free-streaming. 
From Eq.~\eqref{eq:Omegaw}, this is equivalent to a lower bound on $f_a$:
\begin{equation}
   f_a\gtrsim 5\cdot 10^9 \GeV \frac{f_{\mathrm{DM}}^{5/6}}{\mathcal{A}_d^{1/2}C_d^{1/2}}\left(\frac{10^{-6}\eV}{m_a}\right)^\frac12 \, .
\end{equation}
Note that, (a weaker version of) this limit also applies for $f_{\rm DM} \ll 1$ provided we use the Lyman-$\alpha$ EFT constraint shown in Fig.~\ref{fig:Piso}, and the effective cutoff satisfies $k_{\rm dw} \gtrsim 0.05\,{\rm Mpc}^{-1}$, beyond which this limit disappears. This corresponds approximately to $f_{\rm DM} \gtrsim 2 \times 10^{-7}$.
Refining these bounds and studying substructure in this scenario would require a better understanding of the axion abundance and  spectrum generated by domain walls.

\section{Summary and future directions}
\label{sec:conclusions}

We have presented a comprehensive analysis of post-inflationary axions and provided the most stringent cosmological bounds on these to date. Specifically, using large-scale structure observations across a wide range of scales, we constrained the white-noise isocurvature fluctuations in the axion dark matter field. 
Our key result is a new search of such fluctuations based on the ultraviolet luminosity function (UVLF) of galaxies observed by the Hubble Space Telescope (HST) at redshifts $z=4$–$10$, which probes the matter distribution on comoving scales $k\simeq0.5$–$10\,{\rm Mpc}^{-1}$. By comparing HST data with the UVLF predicted from the axion-modified matter power spectrum, we derived new bounds on subdominant axion dark matter fractions in the mass range $m_a\simeq 10^{-26}-10^{-24} \,{\rm eV}$ and decay constants $f_a\simeq10^{14}$–$10^{15}\,{\rm GeV}$ (see Fig.~\ref{fig:money_plot}). We also derived new model-independent limits from the UVLF on generic white-noise power spectra as a function of their momentum cutoff $k_{\rm cut}$ (see Fig.~\ref{fig:Piso}).

The white-noise fluctuation spectrum arises from the decay of axion strings and domain walls, and is modeled using inputs from field-theory simulations. Specifically, the spectrum is peaked at the comoving momentum $k_{\rm wn}=C k_\star$, with $k_\star\propto m_a^{1/2}$ in Eq.\,\eqref{eq:kstar} and $C=\mathcal{O}(10)$. Thus, lighter axions seed fluctuations on larger comoving scales, producing more massive early halos and a correspondingly stronger isocurvature signal. A main result of our work is that, although $C \gg 1$ renders the axions relatively warm—suppressing fluctuation growth above the velocity dispersion Jeans scale $k_J^{\rm eq}$—they can still drive detectable structure formation in the dominant CDM component.

Combining the UVLF constraints with complementary probes (including cosmic microwave background anisotropies, baryon acoustic oscillations, and the Lyman-$\alpha$ forest) provides a consistent picture of the post-inflationary axion scenario across nearly ten orders of magnitude in the axion mass. These data exclude large decay constants, where the string network would induce excessive inhomogeneities, and very low masses, where the white-noise tail would accelerate the formation of early halos and galaxies. Additionally, radio interferometers such as the Square Kilometre Array will be sensitive to the stochastic gravitational-wave background emitted by axion strings, and will thus refine further these limits. 

In this work, we studied enhanced structure formation from the white-noise axion tail in the regime $f_{\rm DM} \ll 1$ where the suppression of perturbations due to velocity dispersion is small. Extending the analysis to $f_{\rm DM} = \mathcal{O}(1)$ would be valuable, but requires assessing the combined impact of free streaming, quantum/classical Jeans suppression, and isocurvature white-tail enhancement using full transfer functions. This is nontrivial, as standard analyses—including N-body simulations and UVLF or Lyman-$\alpha$ modeling—assume normal CDM evolution, although such an extension is possible along the lines of Refs.~\cite{Amin:2025dtd,Amin:2025sla}. In any case, we expect that theoretical uncertainties in the axion relic abundance and overdensity spectrum from strings and domain walls continue to dominate current bounds.

Ongoing and forthcoming datasets—including high-resolution spectroscopy of distant quasars and galaxies (e.g., Lyman-$\alpha$ from DESI~\cite{2013arXiv1308.0847L} and Spec-S5~\cite{Spec-S5:2025uom}) and wide-field surveys such as \textit{Euclid}~\cite{Laureijs:2011gra} and the Rubin Observatory LSST~\cite{Abate:2012za}—will extend the precision and redshift reach of small-scale clustering measurements. Complementary probes from 21\,cm observations~\cite{Munoz:2019hjh,deKruijf:2024voc} and deep-field imaging with JWST~\cite{2023Natur.616..266L,2024Natur.635..311X} will sharpen constraints on the earliest galaxies and the end of the cosmic dark ages. In particular, forthcoming JWST UVLFs, with access to fainter magnitudes and higher redshifts, will enable more stringent tests of the post-inflationary axiverse; as these datasets mature, the constraints presented here can be systematically updated. 

More broadly, high-redshift large-scale structure measurements provide a direct connection between early-universe field dynamics and the observed galaxy distribution, allowing us to probe a wider class of particle-physics models. Ultimately, this leverage offers a unique path to testing dark sectors that communicate with the visible sector only through gravity.

\section*{Acknowledgements}

We thank Mustafa Amin for interesting discussions and contributions at various stages of this project. 
We also thank Andrew~Long, Mehrdad~Mirbabayi, Diego~Redigolo, Nash~Sabti, Giovanni~Villadoro, Ed~Hardy, Wayne~Hu, and Tanvi~Karwal for useful discussions.
The work of M.G. is supported by the Alexander von Humboldt foundation and has been partially funded by the Deutsche Forschungsgemeinschaft under Germany’s Excellence Strategy - EXC 2121 Quantum Universe - 390833306. S.T. was supported by the Swiss National Science Foundation - project n. P5R5PT\_222350, and acknowledges CERN TH Department for hospitality while this research was being carried out. This project has received also funding from the European Union’s Horizon Europe research and innovation programme under the Marie Skłodowska-Curie Staff Exchange grant agreement No 101086085 – ASYMMETRY.  G.V. acknowledges the support of the Eric and Wendy Schmidt AI in Science Fellowship at the University of Chicago, a program of Schmidt Sciences.


\printbibliography

\end{document}